\documentclass[useAMS,usenatbib]{mnras}
\usepackage{epsfig,rotate,graphicx}
\usepackage[fleqn]{amsmath}
\usepackage{subfig}
\usepackage{lscape}
\usepackage{bm}
\usepackage{tikz}
\usepackage{paralist}
\usetikzlibrary{shapes.geometric, arrows}

\hypersetup{draft}

\newcommand{\p}{\partial}

\newcommand{\be}{\begin{equation}}
\newcommand{\ee}{\end{equation}}
\newcommand{\gtrsim}{\;\raisebox{-.8ex}{$\buildrel{\textstyle>}\over\sim$}\;}
\newcommand{\lesssim}{\; \raisebox{-.8ex}{$\buildrel{\textstyle<}\over\sim$}\;}

\newcommand{\avg}[1]{\left\langle #1 \right\rangle}
\newcommand{\avgt}[1]{\left\langle #1 \right\rangle_t}

\newcommand{\rhod}{\rho_\mathrm{d}}
\newcommand{\rhog}{\rho_\mathrm{g}}

\newcommand{\vd}{\bm{v}_\mathrm{d}}
\newcommand{\vg}{\bm{v}_\mathrm{g}}
\newcommand{\tstop}{t_\mathrm{s}}
\newcommand{\fdust}{f_\mathrm{d}}
\newcommand{\OmK}{\Omega_\mathrm{K}}
\newcommand{\OmF}{\Omega_\mathrm{K0}}

\newcommand{\Hgas}{H_\mathrm{g}}
\newcommand{\Hdust}{H_\mathrm{d}}
\newcommand{\hgas}{h_\mathrm{g}}

\newcommand{\Tstop}{T_\mathrm{stop}}
\newcommand{\taus}{\tau_\mathrm{s}}

\newcommand{\Sigd}{\Sigma_\mathrm{d}}
\newcommand{\Sigg}{\Sigma_\mathrm{g}}
\newcommand{\vzmid}{v_\mathrm{z0}}
\newcommand{\teddy}{t_\mathrm{eddy}}
\newcommand{\reyz}{\mathcal{R}_{z\phi}}
\newcommand{\reyzt}{\mathcal{R}_{z\phi\mathrm{,}t}}

\defcitealias{lin17}{LY17}
\defcitealias{lin15}{LY15}
\defcitealias{chen18}{CL18}

\usepackage{array,booktabs,tabularx}
\newcolumntype{R}{>{\centering\arraybackslash}X} 

\title[Dust settling against turbulence]{Dust settling against hydrodynamic turbulence in protoplanetary discs}  
\author[Lin]{Min-Kai Lin\thanks{mklin@asiaa.sinica.edu.tw}
\\  Institute of Astronomy and Astrophysics, Academia Sinica, Taipei 10617, Taiwan
} 

\begin{document}

\maketitle
\begin{abstract}
Enhancing the {  local} dust-to-gas ratio in protoplanetary discs is a necessary first step to planetesimal  formation. In laminar discs, dust settling is an efficient mechanism to raise the dust-to-gas ratio at the disc midplane. 
However, turbulence, if present, can stir and lift 
dust particles, which ultimately hinders planetesimal formation. In this work, we study dust settling in protoplanetary discs {  with} hydrodynamic turbulence sustained by the vertical shear instability. We perform axisymmetric numerical simulations to investigate the effect of turbulence,  particle size, and {  solid abundance or} metallicity on dust settling. 
We highlight the positive role of {  drag forces exerted onto the gas by the dust for} settling to overcome the vertical shear instability. In typical disc models we find particles with a Stokes number $\sim 10^{-3}$ can sediment to $\lesssim 10\%$ of the gas scale-height, provided that $\Sigd/\Sigg\gtrsim 0.02$---0.05, where $\Sigma_\mathrm{d,g}$ are the surface densities in dust and gas, respectively. This coincides with the metallicity condition for {  small particles to undergo clumping via} the streaming instability. 
Super-solar metallicities, at least locally, are thus required for a self-consistent picture of planetesimal formation. Our results also imply that dust rings observed in protoplanetary discs should have smaller scale-heights than dust gaps{ , provided that the metallicity contrast between rings and gaps exceed the corresponding contrast in gas density.} 
\end{abstract}


\section{Introduction}\label{intro}


In the standard `core accretion' scenario, planets are formed by the aggregation of km-sized or larger planetesimals via their mutual gravitational attraction \citep{safronov69,lissauer93,raymond14,helled14}. The precursor to planet formation is therefore planetesimal formation \citep{chiang10}.  
However, solids in protoplanetary discs (PPDs) begin as micron-sized dust. 
These small grains can grow to mm-cm sizes by sticking, but collisional growth beyond this size is limited by bouncing or  fragmentation \citep{blum18}. Other effects, such as turbulence or radial drift, can also hinder particle growth \citep{johansen14}. 

One way to circumvent these growth barriers is to invoke the collective gravitational collapse of a population of dust grains 
\citep{goldreich73}. {  A} dynamical gravitational instability (GI) in dust requires a high dust-to-gas ratios, $\rhod/\rhog\gg 1$, where $\rhod$ and $\rhog$ are the dust and gas mass densities, respectively \citep{chiang10,shi13}. 
{  However, in a newly formed PPD, where dust and gas are well-mixed, we expect $\rhod/\rhog\sim 10^{-2}$ everywhere, as typical of the dust-to-gas mass ratio in the interstellar medium.} 
Thus, to trigger GI for planetesimal formation, dust grains must first be concentrated to sufficiently high volume densities relative to the gas.  

A promising mechanism for concentrating dust is the streaming instability \citep[SI,][]{youdin05a,youdin07b}. The SI naturally arises from the mutual dust-gas drag in rotating discs \citep{jacquet11}, and is driven by the relative drift between dust and gas caused by radial pressure gradients. More recent physical interpretations of the SI have been given by \citet[][hereafter \citetalias{lin17}]{lin17} and \cite{squire17}. 

In principle, the SI can occur {  at} any dust-to-gas ratios, but for small particles 
the SI grows slowly {  under dust-poor conditions ($\rhod\lesssim \rhog$)}. Furthermore, in this limit the SI does not lead to dust clumping \citep{johansen07}. Indeed, numerical simulations show strong clumping, and hence planetesimal formation, only when $\rhod \gtrsim \rhog$ \citep[e.g.][]{johansen09, bai10,bai10c,yang16b,simon16}. How to realise {  this} optimal condition for the SI { , at least locally in a PPD, is} therefore a crucial issue for planet formation.

Several other dust-concentrating mechanisms may act to seed the SI \citep{chiang10, johansen14}. These include dust-trapping by pressure bumps{ , zonal flows,} and vortices 
{  \citep{barge95,johansen09b,dittrich13,lyra13,raettig15,pinilla17}}; gas removal by photoevaporation or disc winds \citep{alexander14,bai16}; other dust-gas drag instabilities \citep{takahashi14,gonzalez17}; radial pile-up of dust \citep{kanagawa17}; and dust settling \citep{nakagawa86,dubrulle95,takeuchi02}. Some of these processes may directly induce GI in dust without going through SI.

In this work we focus on dust settling. This is perhaps the simplest way to increase $\rhod/\rhog$: dust particles sediment towards the disc midplane due to the vertical component of stellar gravity and drag forces from the gas. 
However, dust settling requires the disc to be sufficiently laminar \citep{dubrulle95,youdin07}. Otherwise, turbulence can stir up the particles. Turbulence may arise from the settling process itself, for example via Kelvin-Helmholtz instabilities associated with {  the} differential rotation between the dust and gas layers \citep[KHI,][]{johansen06,barranco09,chiang08,lee10}, or weak SI \citep{johansen07,bai10}. In these cases, $\rhod/\rhog$ may be self-limited and never reach unity.  


Furthermore, there are external sources of turbulence in PPDs. Magnetic {  instabilities} \citep{balbus91} and gaseous  GI \citep{gammie01} can have significant impact on dust dynamics  \citep[e.g.][]{rice04,fromang06,balsara09,gibbons12,zhu15,shi16,riols18,yang18}. 

In addition, there is a plethora of newly (re)discovered, purely hydrodynamic instabilities that drive turbulence \citep{fromang17,klahr18,lyra18}. 
These include: the zombie vortex instability \citep{marcus15,lesur16,umurhan16}, the convective overstability \citep{klahr14,lyra14,latter16}, and the vertical shear instability (VSI, \citealt{nelson13}; \citealt{lin15}, hereafter \citetalias{lin15}; \citealt{barker15}). These instabilities are relevant to magnetically inactive or non-self-gravitating regions of the disc. 
Among them, the VSI is particularly relevant to dust settling.


The VSI taps into the free energy associated with a vertical gradient in the gaseous disc's angular velocity $\Omega$ (\citetalias{lin15}, \citealt{barker15}). Such a vertical shear, $\p_z~\Omega\neq~0$, arises  whenever the gas is baroclinic. This is the case in the outer parts of PPDs where a non-uniform radial temperature profile is set by stellar irradiation \citep{stoll14}. The VSI is characterised by large-scale vertical motions \citep{nelson13} and the associated stress is much stronger than that in the radial direction \citep{stoll17b}. It is then natural to question whether or not dust particle can still settle to the point of triggering rapid SI. 


Recent numerical simulations have begun to study the influence of the VSI on particle dynamics in PPDs \citep{stoll16,flock17}. These studies find that small particles with sizes $\lesssim \text{mm}$ can be efficiently lifted by VSI turbulence. However, these studies model dust as passive particles, i.e. the back-reaction of dust-drag onto the gas is neglected. However, 
this feedback would induce an effective buoyancy force, as the dust adds weight to the mixture but not pressure \citepalias{lin17}. On the other hand, buoyancy is known to be effective 
in stabilising the VSI \citepalias{lin15}. It is thus necessary to include particle back-reaction onto the gas to study dust settling against VSI turbulence.  

In this work, we {  study dust settling against the VSI by applying} \citetalias{lin17}'s thermodynamic model of dusty PPDs, which implicitly includes particle back-reaction. This allows us {  investigate} the effect of the overall solid abundance on dust settling in VSI-active discs. \citetalias{lin17}'s framework was previously applied to study dusty  disc-planet interaction \cite[][hereafter \citetalias{chen18}]{chen18}. Here we adapt \citetalias{chen18}'s numerical models. 

This paper is organised as follows. In \S\ref{model} we briefly review the governing equations of \citetalias{lin17} and \citetalias{chen18}, which forms the basis of this work. We then describe our initial disc models in \S\ref{initial_condition}. Our numerical setup is described in \S\ref{simulations}, along with several diagnostics. We present simulation results in \S\ref{results}. We find that dust settling can overcome the VSI if the local metallicity, $\Sigd/\Sigg$, where $\Sigma_\mathrm{d,g}$ are dust and gas surface densities, respectively, is a few times above the solar value. We interpret and discuss our results {  in \S\ref{discuss}}, including its implication for planetesimal formation {  and observations of PPDs}. We conclude and summarise in \S\ref{summary}. 

\section{Basic equations}\label{model}

We consider an accretion disc comprised of gas and dust orbiting a
central star of mass $M_*$. We neglect disc self-gravity, magnetic fields, and
viscosity. Cylindrical $(R,\phi, z)$ and spherical 
$(r,\theta,\phi)$ co-ordinates are centred on the star. In the discussions below, we use subscript `0' to denote evaluation at $(R,z)=(R_0, 0)$, where $R_0$ is a reference radius. The Keplerian frequency is $\OmK(R)\equiv\sqrt{GM_*/R^3}$, where $G$ is the gravitational constant. We adopt a rotating frame with angular frequency  $\Omega_\mathrm{K0}\hat{\bm{z}}$. 

The gas phase has density, pressure, and velocity fields $(\rhog, P, 
\bm{v}_\mathrm{g})$. We model a single species of dust particles as a second, pressureless fluid with mass density and velocity $(\rhod,\bm{v}_\mathrm{d})$. The fluid approximation for dust is valid for sufficiently small particles \citep{jacquet11}. 

We also define the total mass density 
\begin{align}
\rho \equiv \rhog + \rhod,
\end{align}
the centre-of-mass velocity 
\begin{align}
\bm{v} \equiv \frac{ \rhod\vd + \rhog\vg }{\rho},
\end{align}
and an effective energy 
\begin{align}
 E \equiv \frac{P}{\gamma-1} + \frac{1}{2}\rho|\bm{v}|^2 + \rho\Phi, 
\end{align}
where {  $\gamma$ is a constant defined below and} $\Phi$ is a time-independent gravitational potential ($\p_t\Phi = 0$). {  Our} numerical simulations evolve the hydrodynamic set $(\rho,\bm{v}, E)$, see \S\ref{evol_eqn}.

\subsection{Polytropic gas}

We consider a locally polytropic gas with 
\begin{align}
 P = c_s^2(R)\rho_\mathrm{g0}\left(\frac{\rhog}{\rho_\mathrm{g0}}\right)^\gamma \equiv K(R)\rhog^\gamma, \label{poly_eos}
\end{align}
where $c_s(R)$ is a prescribed sound-speed profile and $\gamma$ is the constant polytropic index. We take $\gamma=1.001$ so the gas is nearly isothermal {  in its thermodynamic property}. We set
\begin{align} 
  c_s^2(R) = c_{s0}^2 \left(\frac{R}{R_0}\right)^{-q}, \label{temp_profile}
\end{align}
where $q$ is the constant temperature profile index. {  Such a temperature profile is said to be locally or vertically isothermal.} 
We also define the pressure scale height $\Hgas$ 
\begin{align}
\Hgas \equiv \frac{c_s}{\OmK},
\end{align}
and the gas disc aspect ratio is $\hgas \equiv \Hgas/R$. 

\subsection{Tightly coupled dust}

The dust fluid interacts with the gas via a drag force characterised by {  the particle} stopping time $\taus$. We consider particles in the Epstein regime with { 
$  \taus = \rho_\bullet s /\rhog c_s $}
\citep{weidenschilling77}, where $\rho_\bullet$, $s$ is the
internal density and radius of a grain, respectively. 

{  However, in the one-fluid framework adopted below, it is convenient to define the \emph{relative} stopping time $\tstop = \taus\rhog/\rho$ \citep{laibe14}.} 
This is the characteristic decay timescale for the relative drift between dust and gas, $\left|\bm{v}_\mathrm{d}-\bm{v}_\mathrm{g}\right|$, due to the mutual drag forces. {  The Stokes number can be defined as $\tstop\OmK.$} 

We parameterise {  $\tstop$ via a 
dimensionless stopping time $\Tstop$} 
such that 
\begin{align} 
  \tstop = \frac{\rho_\bullet s}{\rho c_s} \equiv 
  \frac{\rho_0c_{s0}}{\rho
    c_s}\frac{\Tstop}{\Omega_\mathrm{K0}}\label{drag_law}.    
\end{align}
{  Thus $\Tstop = \Omega_\mathrm{K0}\rho_\bullet s / \rho_0 c_{s0}$ is also the Stokes number for a particle of a given size and internal density at the reference point. For convenience we will also refer to $\Tstop$ as the particle size. 
}


In addition, we consider small particles with  
\begin{align}
  \tstop\OmK \ll 1.
\end{align} 
In this limit dust particles are tightly, but not necessarily perfectly, coupled to the gas; and one can relate dust and gas velocities by the 
\emph{terminal velocity 
approximation}:
\begin{align} 
\vd = \vg + \tstop \frac{\nabla P}{\rhog}\label{term_approx}
\end{align}
\citep[][\citetalias{lin17}]{youdin05a,jacquet11,price15}. {  Note that a differential Coriolis force has been neglected in Eq. \ref{term_approx} as it is of order $\tstop^2$ \citep[see, e.g.][]{laibe14b}.} Eq. \ref{term_approx}  reflects the fact that particles tend to drift towards pressure maxima 
\citep{whipple72,weidenschilling77}. 


\subsection{Evolutionary equations}\label{evol_eqn}
With the above definitions and approximations, \citetalias{lin17} and \citetalias{chen18}, building upon \cite{laibe14}, showed that the dusty gas disc can be described by the following evolutionary equations:
\begin{align}
	 & \frac{\p\rho}{\p t} + \nabla\cdot\left(\rho\bm{v}\right)= 0,\label{masseq}\\
  	&\frac{\p\bm{v}}{\p t} + \bm{v}\cdot\nabla\bm{v} = 
  -\frac{1}{\rho}\nabla P - \nabla \Phi - 	2\OmF\bm{\hat{z}}\times\bm{v},\label{momeq}\\ 
& \frac{\p E}{\p t} + \nabla\cdot\left[(E+P)\bm{v}\right]  =  
  \frac{P}{\gamma-1}\bm{v}\cdot\nabla\ln{c_s^2} \notag \\
&  \phantom{\frac{\p E}{\p t} + \nabla\cdot\left[(E+P)\bm{v}\right]=}
   + \frac{\gamma P}{\rhog(\gamma
    -1)}\nabla\cdot\left(\fdust\tstop\nabla P\right), \label{dusteq}
\end{align}
where the dust fraction $\fdust \equiv  \rhod/\rho$ can be obtained from the equation of state (Eq. \ref{poly_eos})
\begin{align}
  \fdust = 1 - \frac{1}{\rho}\left(\frac{P}{K}\right)^{1/\gamma}, \label{fdust_replace}
\end{align}
and $\rhog=(1-\fdust)\rho$. {  We define} the dust-to-gas ratio $\epsilon \equiv \rhod/\rhog =
\fdust/\left(1-\fdust\right)$. {  A similar one-fluid model was also adopted by \cite{fromang06} to simulate dusty, magnetised discs.}

In the rotating frame the total gravitational potential is      
\begin{align} 
\Phi(R,\phi,z) =& \Phi_*(R,z) -
\frac{1}{2}R^2\OmF^2. 
\end{align}
Here, 
\begin{align}
  \Phi_*(R,z)  = - R^2\OmK^2 \left(1 - 
    \frac{z^2}{2R^2}\right).\label{thin_pot} 
\end{align}
is the stellar potential in the thin-disc approximation ($|z|\ll R$). We adopt this approximate form in order specify initial conditions analytically. 

Eqs. \ref{masseq}---\ref{dusteq} provides a hydrodynamic description of the dusty gas.  
The dust density is recovered from the equation of state (\ref{poly_eos}), and the dust velocity from the terminal velocity approximation ($\bm{v}_\mathrm{d} = \bm{v} + \tstop\nabla P/\rho$). Eq. \ref{masseq}---\ref{momeq} physically corresponds to mass and momentum conservation for the mixture.

However, Eq. \ref{dusteq} is not a physical energy equation. It simply stems from dust mass conservation, the equation of state, and the terminal velocity approximation (\citetalias{lin17}; \citetalias{chen18}). The term $\propto \nabla c_s^2$ on the right-hand-side models the rapid cooling that enforces a locally (vertically) isothermal {  temperature profile}, and is the term responsible for the VSI. The second term on the right-hand-side models dust-gas decoupling. 

\subsection{Dusty entropy and buoyancy}\label{dusty_nz}
Describing the dusty gas in terms of the hydrodynamic variables $(\rho, \bm{v}, P)$ allows us to 
define the (dimensionless) effective entropy of the system as 
\begin{align} 
S \equiv \ln{\frac{P^{1/\gamma}}{\rho}} = \ln{c_s^{2/\gamma}} - \ln{(1+\epsilon)} + \mathrm{const.}
\end{align}
\citepalias{lin17}. A non-uniform entropy profile produces buoyancy forces. For this work the relevant quantity is the {  squared} vertical buoyancy frequency 
\begin{align}
N_z^2 \equiv -\frac{1}{\rho}\frac{\p P}{\p z}\frac{\p S}{\p z} = \frac{1}{1+\epsilon}\frac{1}{\rho}\frac{\p P}{\p z}\frac{\p \epsilon}{\p z}, 
\label{dusty_buoyancy}
\end{align}
where the second equality applies to the vertically isothermal {  temperature profiles} we consider. 

The notion of a dust-induced buoyancy is most applicable to dust perfectly coupled to the gas. In this limit, dust particles add to the mixture's inertia but not the pressure. Dust therefore `weighs down' the fluid. Hence there is an associated buoyancy force. Settled dust provides a stabilising force against vertical motions of the VSI since $N_z^2>0$ in that case {(see also \S\ref{buoyant_stabilisation}).}

However, settling in the first place requires some decoupling between dust and gas. This diminishes the effective buoyancy as dust can slip past the gas. The definition above should thus be interpreted as the maximum buoyancy induced by dust. 


\section{Disc models}\label{initial_condition}

We initialise our discs by first specifying the dust-to-gas ratio distribution 
\begin{align}
 \epsilon(R,z) = \epsilon_\mathrm{mid}(R) \times \exp{\left( -
  \frac{z^2}{2H_\epsilon^2}\right)}.\label{dg3D}
\end{align}  
Here and below subscript `mid' indicates the midplane value. The profile $\epsilon_\mathrm{mid}(R)$ is chosen to minimise radial disc evolution, as described in Appendix B of \citetalias{chen18}. The characteristic thickness in the dust-to-gas ratio, $H_\epsilon$, is defined via 
\begin{align}
 \frac{1}{H_\epsilon^2} \equiv \frac{1}{H_\mathrm{d}^2} -  \frac{1}{\Hgas^2}. \label{Hepsilon}
\end{align}
{  Here the dust layer thickness $H_\mathrm{d}$ is set to $0.99\Hgas$ so that initially the}  
 dust is vertically well-mixed with the gas ($H_\epsilon\gg 1$). {  For $t>0$ we fit the vertical dust density distribution to a Gaussian profile to obtain $\Hdust$ (see \S\ref{diagnostics}).}

We scale the dust-to-gas ratio by specifying the metallicity $Z$ at the reference radius   
\begin{align} 
Z \equiv \left.\frac{\Sigd}{\Sigg}\right|_0 = \left.\epsilon_\mathrm{mid}\frac{H_\mathrm{d}}{\Hgas}\right|_0 \label{Z_def}
\end{align}
\citep{johansen14}. Recall $\Sigma_\mathrm{d,g}$ are the surface densities in dust and gas,  respectively.

We obtain the disc structure by assuming dynamical equilibrium and 
axisymmetry:    
\begin{align}
  R\Omega^2 &= R\OmK^2\left(1 - \frac{3}{2}\frac{z^2}{R^2}\right) + \frac{1}{\rho}\frac{\p P}{\p
    R},\label{rad_eqm3D}\\ 
  0 & = z\OmK^2 + \frac{1}{\rho}\frac{\p P}{\p z}. \label{vert_eqm3D}
\end{align}
Here $\Omega\equiv v_\phi/R + \Omega_\mathrm{K0} $ is the angular velocity in the inertial frame. 
For simplicity we set $\gamma=1$ to integrate Eq. \ref{vert_eqm3D}, making use of Eq. \ref{dg3D}, to obtain 
\begin{align}
  \rhog  = \rho_\mathrm{g,mid}\exp{\left\{-\frac{z^2}{2\Hgas^2} -
  \epsilon_\mathrm{mid}\frac{H_\epsilon^2}{\Hgas^2}\left[ 1-
  \exp{\left(-\frac{z^2}{2H_\epsilon^2}\right)}\right]\right\}}.
\end{align}    
We set the midplane gas density to 
\begin{align} 
  \rho_\mathrm{g,mid}(R) = \frac{\Sigma_\mathrm{g}(R)}{\sqrt{2\pi}\Hgas(R)},
\end{align}
with gas surface density 
\begin{align} 
  \Sigma_\mathrm{g}(R)  =
  \Sigma_\mathrm{g0}\left(\frac{R}{R_0}\right)^{-p}, \label{surfden}
\end{align}
and we fix $p=3/2$. The total density is $\rho~=~\rhog(1+~\epsilon)$. 
{  Finally, the} disc orbital frequency is obtained from Eq. \ref{rad_eqm3D},
\begin{align}
&\Omega(R,z) = \OmK\left[1  - \frac{3}{2}\frac{z^2}{R^2} + 
    \frac{P}{\rho\left(R\OmK\right)^2}\left(\gamma
  \frac{\p\ln{\rho_\mathrm{g}}}{\p\ln{R}} - q\right)\right]^{1/2}.
\end{align}

\subsection{Vertical shear}

The disc rotation possesses vertical shear. This can be seen directly from Eq. \ref{rad_eqm3D}---\ref{vert_eqm3D}:    
\begin{align}
 &R\frac{\p \Omega^2}{\p z}  =
  \frac{c_s^2(R)}{\left(1+\epsilon\right)^2}\left\{
  \frac{\p\epsilon}{\p R}\frac{\p\ln{\rhog}}{\p z}
 -\frac{\p\epsilon}{\p z}\frac{\p\ln{P}}{\p R}\right.\notag\\
  &\phantom{ R\frac{\p \Omega^2}{\p z}  =
    \frac{c_s^2(R)}{\left(1+\epsilon\right)^2}\left\{\right\} }
  \left. -\frac{q}{R} \left(1+\epsilon\right)\frac{\p\ln{\rhog}}{\p z}
  \right\}\label{vert_shear}
\end{align}
Vertical shear can thus arise from a non-uniform dust-to-gas ratio distributions (the first two terms), and from the imposed radial temperature gradient (the last term). However, \citetalias{lin17} showed that gradients in $\epsilon$ generally do not lead to axisymmetric instabilities in PPDs, unless sharp radial edges are present. For axisymmetric models, as will be adopted below, dust settling in radially smooth discs only contributes to stabilisation via vertical buoyancy forces \citep[cf. the non-axisymmetric KHI, see][]{chiang08}. 
The VSI in our simulations are associated with the imposed radial temperature gradient.     

 \subsection{Buoyant stabilisation}\label{buoyant_stabilisation}
We expect the VSI to be affected by buoyancy forces when the buoyancy frequency becomes comparable to the vertical shear rate such that 
\begin{align}
    N_z^2 \gtrsim \left|R\frac{\p\Omega^2}{\p z}\right|_\mathrm{temp}
    \equiv \frac{|q|c_s^2}{(1+\epsilon)R}\left|\frac{\p\ln{\rhog}}{\p z}\right|\label{vshear_Tgrad}
\end{align}
\citepalias{lin15}, where the right-hand-side is the last term in Eq. \ref{vert_shear}. This is the only destabilising effect in axisymmetric discs in the limit of vanishing particle size \citepalias{lin17}. Evaluating this inequality using Eq. \ref{dusty_buoyancy} gives 
\begin{align}
    \frac{1}{\left(1+\epsilon\right)}\left|\frac{\p\epsilon}{\p z}\right| \gtrsim \frac{|q|}{R}.
\end{align}
For the $\epsilon$ profiles we adopt (Eq. \ref{dg3D}), we have $\left|\p_z\epsilon\right| \sim ~\epsilon_\mathrm{mid}/H_\epsilon$. Then, considering the fiducial radius without loss of generality, we expect dust-induced buoyancy becomes important for midplane dust-to-gas ratios 
\begin{align}
\epsilon_0\gtrsim \hgas |q| \frac{\Hdust}{\Hgas},
\end{align}
for somewhat settled dust ($\Hdust\lesssim \Hgas$), where the definition Eq. \ref{Hepsilon} was used. Thus we can expect dust-induced buoyant stabilisation of the VSI even when $\epsilon_0\sim \hgas |q| \ll 1$. 
\citetalias{lin17} give a similar estimate and confirm it with linear stability calculations (see their sections 6.1---6.2). 

The VSI is sensitive to dust-loading because the temperature-induced, destabilising vertical shear is weak, so even a small amount of dust{ -induced buoyancy} can stabilise it (\citealt{nelson13};  \citetalias{lin15}; \citetalias{lin17}). We give a numerical example of this effect in Appendix \ref{nofeedback}. 

\section{Numerical simulations}\label{simulations}

We carry out global, axisymmetric numerical simulations using the \textsc{pluto} hydrodynamics code \citep{mignone07,mignone12}, modified to evolve the dust-free model equations \ref{masseq}---\ref{dusteq}, as described in \citetalias{chen18}. We configure \textsc{pluto} with piecewise linear reconstruction 
, the HLLC Riemann solver, and second order Runge-Kutta time integration. {  We found this default setup to be robust\footnote{Test simulations with piecewise parabolic interpolation, a Roe solver, and third order Runge-Kutta time integration produced numerical artefacts near the vertical boundaries.}, but its diffusive nature likely suppresses the streaming instability. }

We adopt spherical co-ordinates with $r\in[0.6,1.4]R_0$ and a 
polar angle range such that $\tan{\left(\pi/2 - \theta\right)}\in[-2,2 ]h_\mathrm{g0}$, i.e. two gas scale-heights above and below the disc midplane at the reference radius. 
We fix $h_\mathrm{g0} = 0.05$. The standard resolution is $N_r \times N_\theta = 1920\times 480$; with logarithmic and uniform spacing in $r$ and $\theta$, respectively. This corresponds to a resolution of approximately 100 cells per $\Hgas$ at $R=R_0$. We impose reflective boundary conditions in $\theta$, and set radial boundary ghost zones to their initial values. 

\subsection{Physical parameters}


Our fiducial parameter values are $q=1$, $\Tstop=10^{-3}$, and $Z=10^{-2}$. As shown below, {  VSI turbulence prevents dust settling for this setup}. We are interested in how these parameters can be varied to enable settling: 

\begin{enumerate}
\item We vary the imposed radial temperature gradient $q~\in~[0,1]$. Smaller $|q|$ decreases the strength of the VSI turbulence and is expected to help settling. 
\item We vary the stopping time $\Tstop\in[10^{-3},10^{-2}]$. Larger $\Tstop$, corresponding to larger particles, is expected to settle faster and avoid stirring by the VSI, at least initially.    
\item We vary the overall metallicity $Z\in[10^{-2},10^{-1}]$. Increasing $Z$ is expected to stabilise the VSI and favour settling. 
\end{enumerate}

For comparison, the oft-used Minimum Mass Solar Nebula (MMSN) has a shallower temperature profile with $q~=~0.43$ and is slightly more dusty at $Z=0.015$ \citep{chiang10} than our fiducial setup. Epstein drag stopping times in MMSN-like disc models vary as 
\begin{align}
\Tstop \simeq \frac{10^{-3}}{F}\left( \frac{s}{100\mu\mathrm{m}} \right)
\left(\frac{\rho_\bullet}{\mathrm{g}\mathrm{cm}^{-3}}\right)\left(\frac{R_0}{20\mathrm{au}}
\right)^{3/2},\label{tstop_est}
\end{align}
where $F$ is the mass scale relative to the MMSN. At fixed $\Tstop$, smaller particles reside at larger distances. In a $F=1$ disc and taking $\rho_\bullet = 1\,\mathrm{gcm}^{-3}$, we find $\Tstop=10^{-3}$ corresponds to particle sizes from $\sim 280\mu\mathrm{m}$ at $10$au to $\sim 9\mu\mathrm{m}$ at $100$au; or between $\sim 3\mathrm{mm}$ and $90\mu\mathrm{m}$ for $\Tstop=10^{-2}$.

\subsection{Diagnostics}\label{diagnostics}

We analyse our simulation results using the diagnostics {  listed below.} 
{  We denote mean value of a quantity $Q$ by $\avg{Q}$. Unless otherwise stated, the spatial average is taken over the shell $\left|r - R_0\right|\leq H_\mathrm{g0}$. We also use $\mathrm{rms}(Q) $ or $ \sqrt{\avg{Q^2}}$ to denote the root-mean-squared value of $Q$. Time-averaged values are denoted with an additional subscript $t$. Time is reported in units of the orbital period at $R_0$ and is denoted by $P_0$. }


\begin{itemize}
	\item The dust scale-height $\Hdust$ is obtained by fitting the dust density with   
    	\begin{align}
			\rhod(R,z) = \rhod(R,0)\exp{\left(-\frac{z^2}{2\Hdust^2}\right)},
       	\end{align}
        where $\rhod(R,0)$ is the mid-plane dust density. {  We then normalise $\Hdust$ by $\Hgas$}. 
        \item 	The midplane dust-to-gas ratio $\epsilon_0\equiv \left(\rhod/\rhog\right)_{0}$. 
        We measure both average and maximum values within the sampling region.  
        \item {  The midplane vertical velocity $\vzmid$}.  
        \item Following \cite{stoll17b}, we measure the {  Reynolds} stresses associated with vertical motions via 
	\begin{align}	
R_{z\phi} \equiv \rho v_z \Delta v_\phi,    \label{alpha_z_def}
\end{align}
where $\Delta$ denotes the deviation from the initial value. {  We then define the normalised vertical stress 
\begin{align}
\reyz \equiv \frac{\mathrm{rms}(R_{z\phi})}{c_{s0}^2\avg{\rho_\mathrm{i}}}
\end{align}
as a dimensionless measure of the VSI strength, where $\rho_\mathrm{i}$ is the initial density field.
{ Note that in time evolution plots below, the average is also taken over the height of the disc.}
} 
\end{itemize}


\section{Results}\label{results}


\subsection{Lifting dust particles by the VSI}

We first illustrate the effect of VSI turbulence on dust settling by varying the radial temperature gradient via the power-law index $q$. Here, we fix the particle size and metallicity such that $\Tstop = 10^{-3}$ and $Z = 0.01$, respectively. Larger $|q|$ leads to stronger vertical shear and hence more vigorous VSI turbulence. 

\begin{figure}
\includegraphics[width=\linewidth]{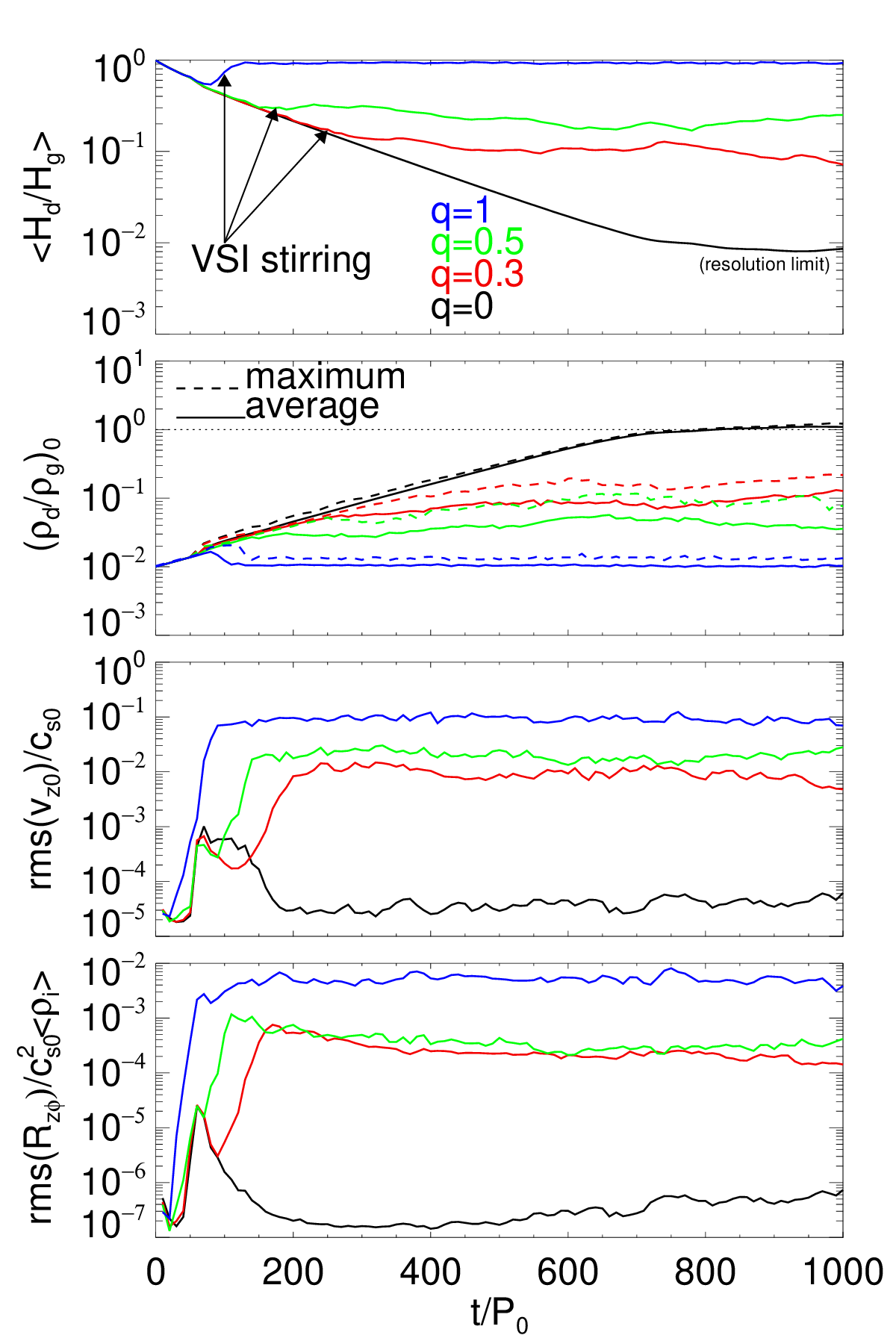}
\caption{
{  Top to bottom: evolution of dust-scale height, midplane dust-to-gas ratios, midplane vertical velocity, and vertical stress parameter;} for different radial temperature gradients. {  The dotted horizontal line in the second panel indicates the critical dust-to-gas ratio, above which the streaming instability is expected to operate efficiently.}
The particle size and metallicity is fixed to $\Tstop=10^{-3}$ and  $Z=0.01$, respectively. 
\label{Hdust_plot_varq}
}
\end{figure}

\begin{figure*}
\includegraphics[scale=.17,clip=true,trim=0cm 4cm 0cm 2cm]{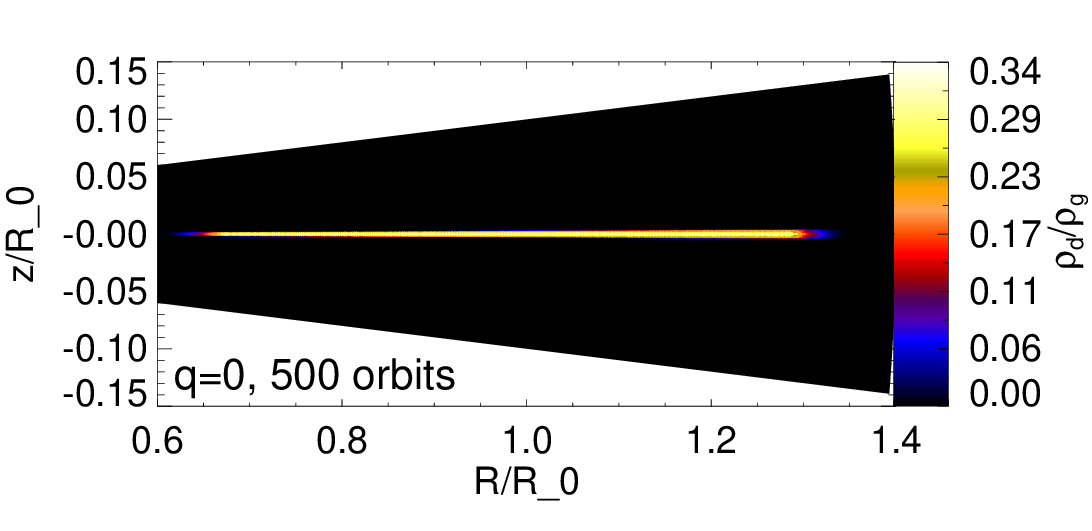}\includegraphics[scale=.17,clip=true,trim=5.5cm 4cm 0cm 2cm]{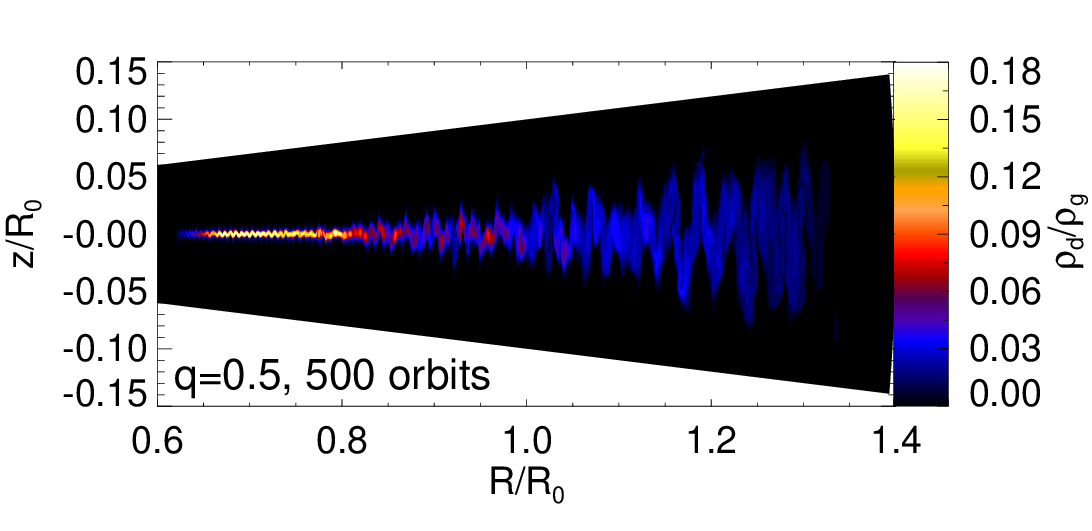}\includegraphics[scale=.17,clip=true,trim=5.5cm 4cm 0cm 2cm]{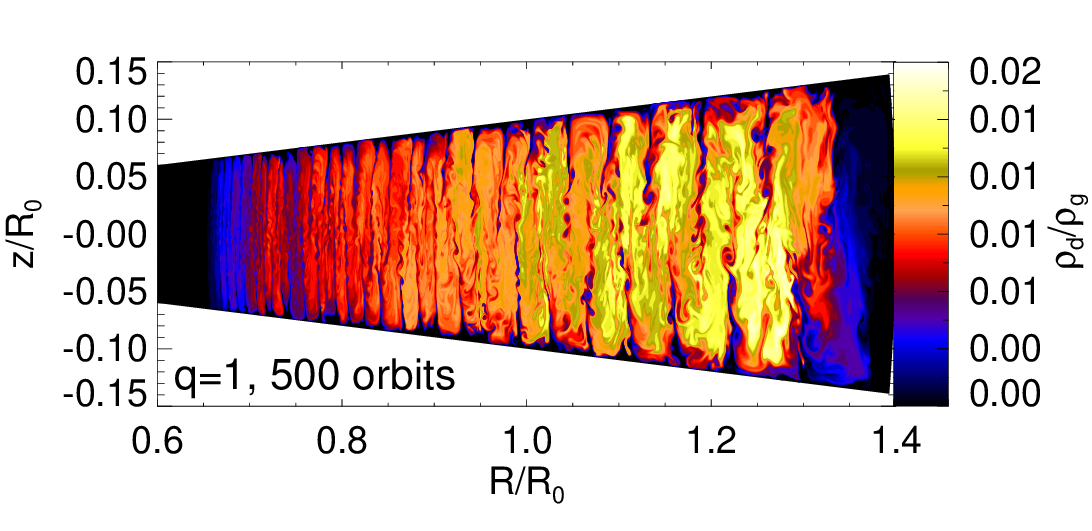}\\
 \includegraphics[scale=.17,clip=true,trim=0cm 1cm 0cm 2cm]{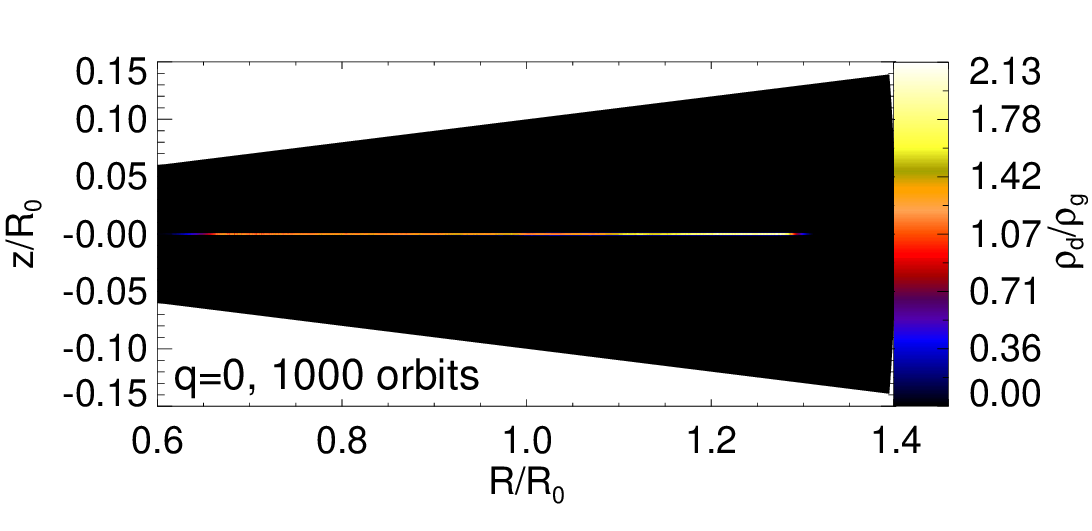}\includegraphics[scale=.17,clip=true,trim=5.5cm 1cm 0cm 2cm]{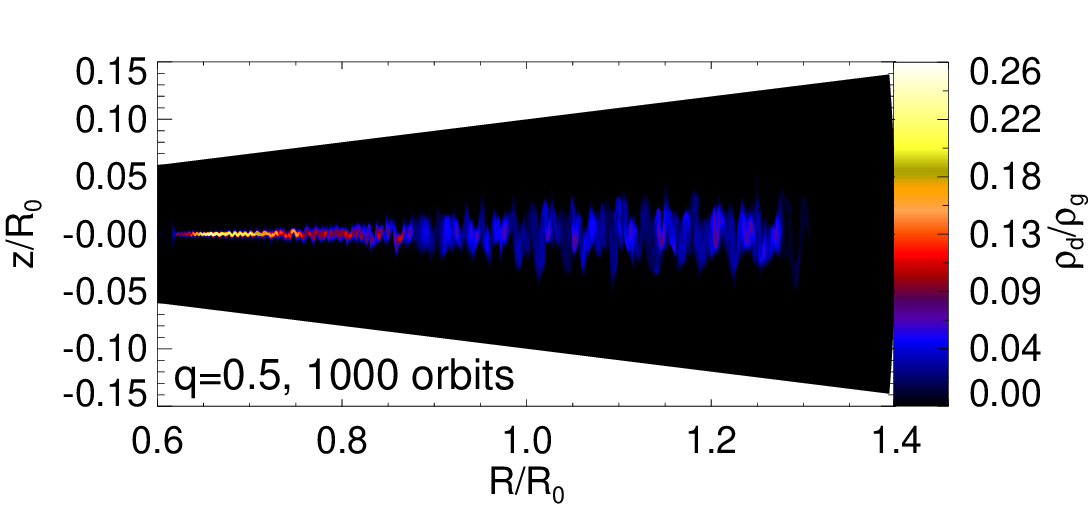}\includegraphics[scale=.17,clip=true,trim=5.5cm 1cm 0cm 2cm]{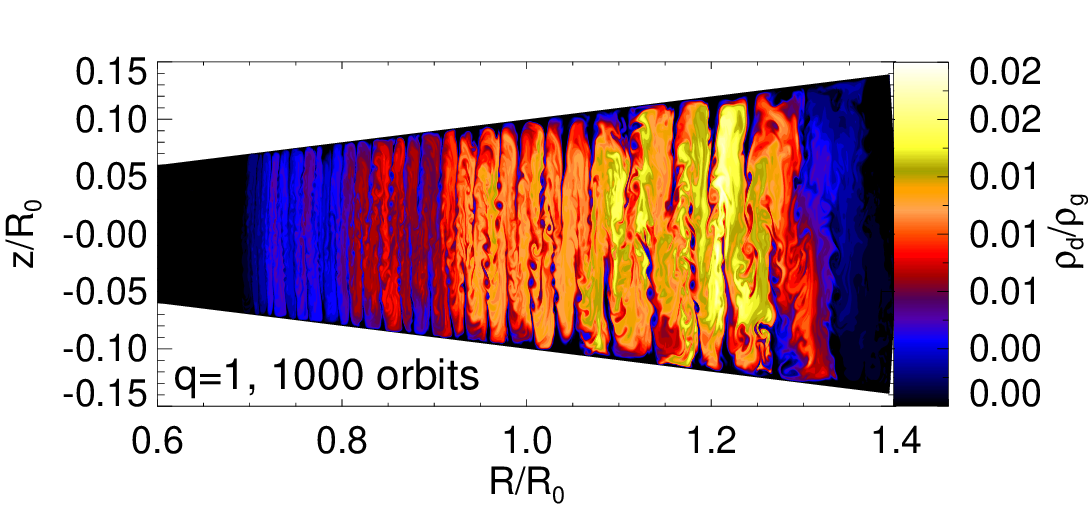}
\caption{Effect of VSI turbulence on dust settling. The dust-to-gas ratio at $t=500P_0$ (top) and $t=1000P_0$ (bottom) are shown for discs with radial temperature profile indices $q=0$ (left, no VSI), $q=0.5$ (middle, moderate VSI), and $q=1$ (right, strong VSI). 
\label{varq_dgratio_compare}} 
\end{figure*}

Fig. \ref{Hdust_plot_varq} shows the time evolution of the {  average} dust scale-height {  normalised by the gas scale-height, $\avg{\Hdust/\Hgas}$}; {  the average and maximum} midplane dust-to-gas ratio,  $\avg{\epsilon_0}$ and $\mathrm{max}(\epsilon_0)$, respectively; {  the root-mean-squared midplane vertical velocity, $\sqrt{\avg{\vzmid^2}}$;} and the {  dimensionless vertical stress, $\reyz$;} for $q=0$, $0.3$, $0.5$, and $1$. Fig.  \ref{varq_dgratio_compare} compares several snapshots of the dust-to-gas ratio from these runs. It is evident that VSI turbulence hinders dust settling.   

For $q=0$ the disc is strictly isothermal and does not develop the VSI. In this laminar disc, dust settling proceeds until the resolution limit ($\simeq 10^{-2}\Hgas$) and $\epsilon_0\sim 1$. Introducing moderate VSI turbulence with $q=0.5$ prevents the dust from settling below $\sim 0.1\Hgas$. Consequently $\epsilon_0$ only increases to $O(0.1)$. 

\begin{figure}
\includegraphics[width=\linewidth]{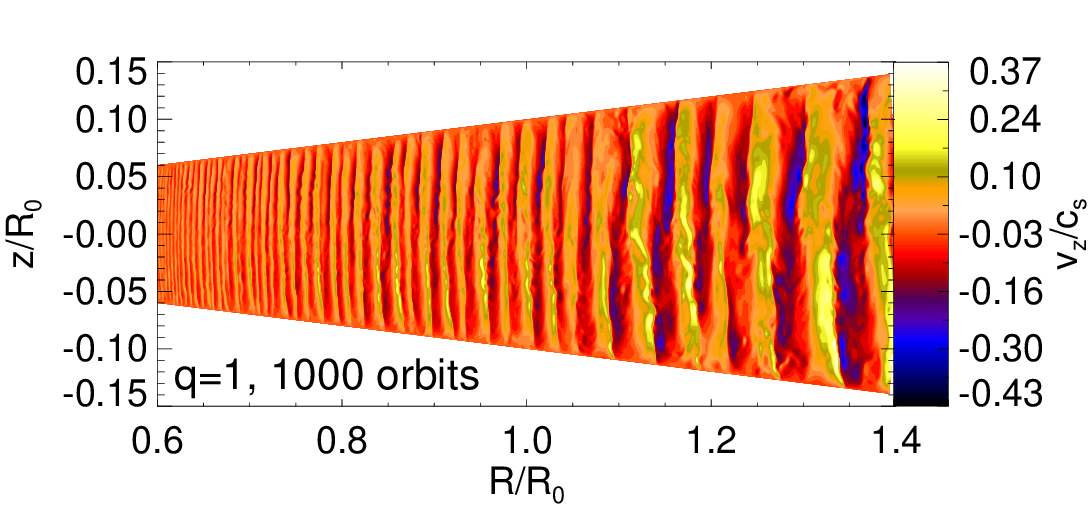} 
\caption{Vertical velocity at the end of the fiducial simulation with $(q,\Tstop,Z)=(1,10^{-3},10^{-2})$. The corresponding dust-to-gas ratio is shown in the {  bottom right} panel of Fig. \protect\ref{varq_dgratio_compare}. 
\label{q1_final_vz}
}
\end{figure}

In the fully turbulent disc with $q=1$  we find no signs of dust settling over the simulation timescale.  {Fig. \ref{varq_dgratio_compare} show in this case that} dust remains well-mixed with the gas throughout the vertical disc column with $\rhod/\rhog \sim 10^{-2}$, {  with undulations near the surface.}  The banded structure in $\rhod/\rhog$ with characteristic separation $\sim\Hgas/2$ 
is caused by the advection of dust particles by the large-scale up/down VSI motions{ , which have no vertical structure}. This is shown in Fig. \ref{q1_final_vz}. The average vertical {  stress} $\reyz \simeq 5\times 10^{-3}$ is consistent with \cite{stoll17b}, who found  $\reyz\lesssim 0.01$ for $|z|\leq 2\Hgas$. 


\begin{figure}
\includegraphics[width=\linewidth]{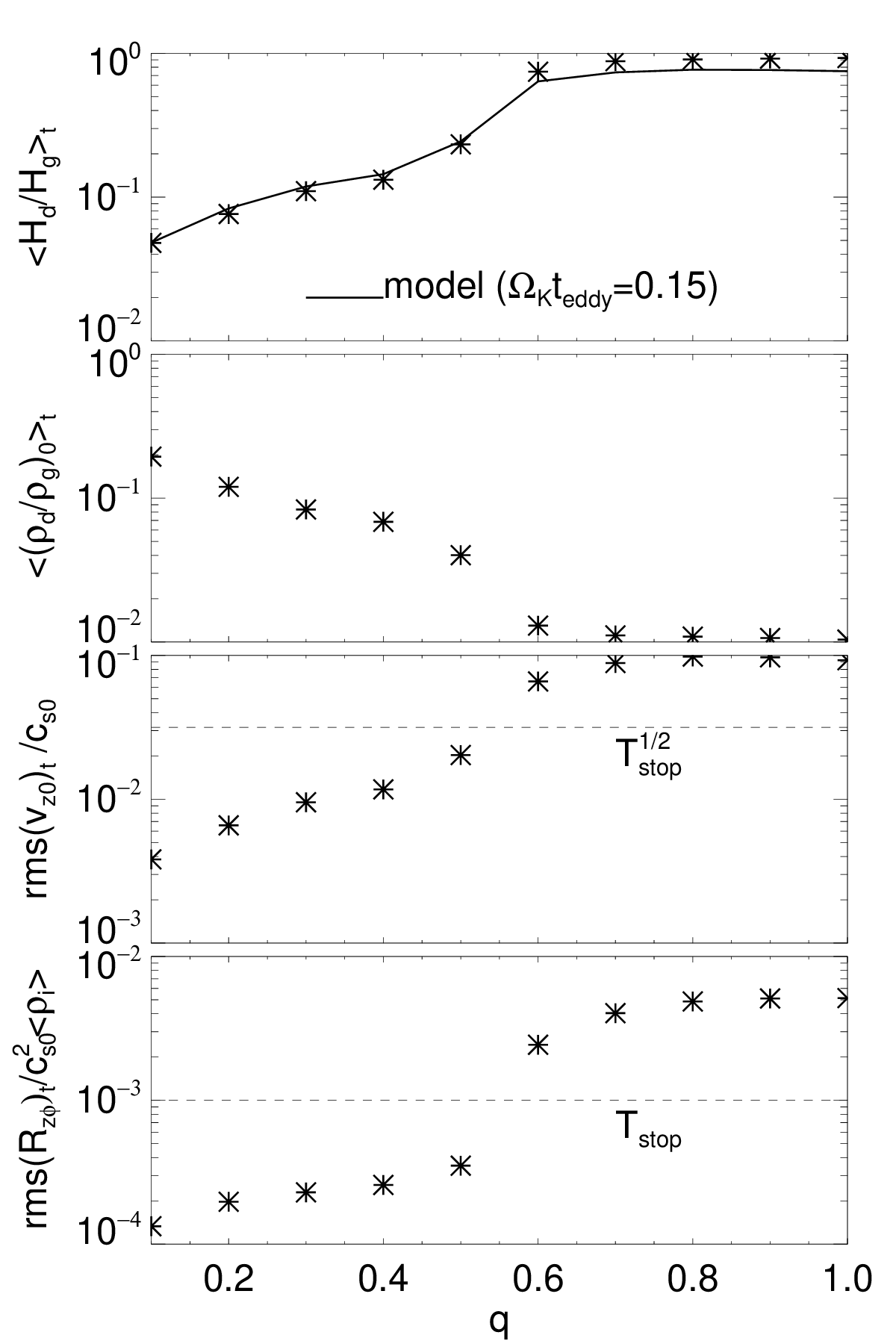}
\caption{
{  Top to bottom: time-averaged dust scale-height, midplane dust-to-gas ratio, midplane vertical velocity, and vertical turbulence strength;} as a function of the radial temperature gradient. {  The solid line in the top panel is obtained from Eq. \ref{hdust_theory}.} 
The particle size and metallicity is fixed to  $\Tstop=10^{-3}$ and $Z=0.01$, respectively. 
\label{varq_Hdust_final}
}
\end{figure}

In Fig. \ref{varq_Hdust_final} we 
{  compare the time-averaged dust distribution and turbulence properties 
} 
from a suite of runs with $q\in[0.1,1]$. {  The averaging period is taken between when the system enters a quasi-steady state and the end of the simulation.} We omit the $q=0$ run here since in that case there is no physical steady state. We find {  $\avg{\Hdust/\Hgas}_t$ increases} rapidly with $q$, but for $q\gtrsim0.5$ the dust becomes well-mixed with $\avg{\Hdust/\Hgas}_t\simeq 1$ and $\avgt{\epsilon_0}\simeq 0.01$ ($=Z$ here). 
{  Notice the well-mixed and settled cases are separated by $\sqrt{\avg{\vzmid^2}}_t\sim \sqrt{\Tstop} c_{s0}$ or $\reyzt \sim \Tstop$. The former dependence is consistent with theoretical dust settling models, described below. }  
As $\avgt{\vzmid^2}$ and $\reyzt$ increases with $|q|$, this suggest that given a fixed particle size there exists a critical temperature gradient, $|q|_*$, below which dust-settling is allowed. However, as shown below, {  vertical stresses and velocities} also depends on particle size and metallicity. 


For comparison, we also plot in Fig. \ref{varq_Hdust_final} the expected dust scale-height
{  based on \cite{dubrulle95}'s model,
\begin{align}
    \frac{H_\mathrm{d,model}}{\Hgas} = \left[1 + \frac{c_{s0}^2\avgt{\OmK\tau_\mathrm{s}}}{\avgt{\vzmid^2}\left(\OmK t_\mathrm{eddy}\right)}\right]^{-1/2}\label{hdust_theory}
\end{align}
\citep[see also][]{youdin07,zhu15}, where the particle stopping time $\taus$ is measured at the disc midplane, and we regard the eddy timescale $t_\mathrm{eddy}$ as a model parameter for simplicity. Setting $\OmK \teddy=0.15$ gives a good match to the measured $\avgt{\Hdust/\Hgas}$ in the settled cases, while it somewhat underestimates the well-mixed cases. This value of $\teddy$ is similar to that found by \cite{stoll16}. 

}

\subsection{Effect of particle size}

We now examine the effect of particle size, which is parameterised by $\Tstop$ in our simulations. Here we fix $q=1$ and $Z=0.01$. 

\begin{figure}
\includegraphics[width=\linewidth]{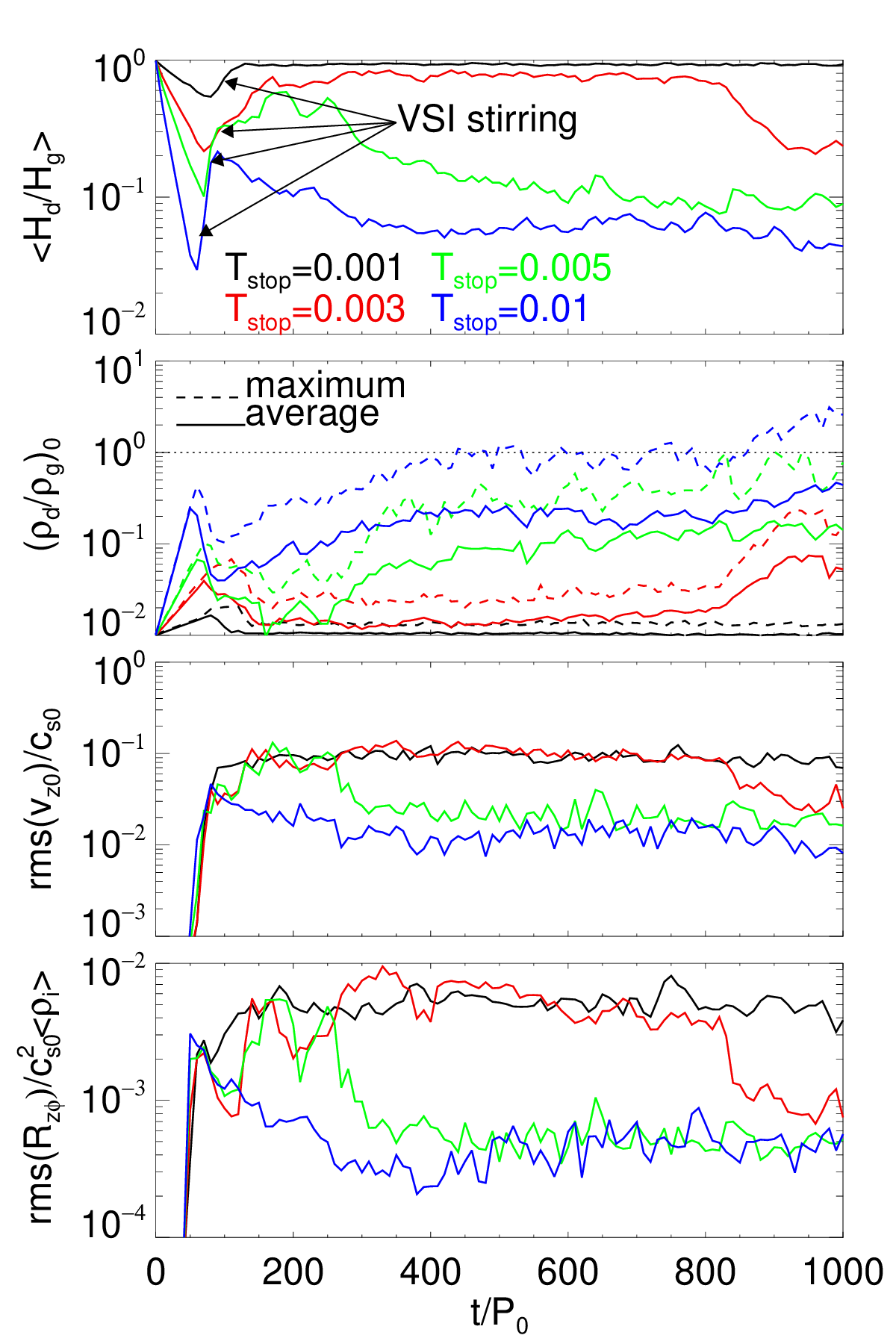}
\caption{
{  
Top to bottom: evolution of dust-scale height, midplane dust-to-gas ratios, midplane vertical velocity, and vertical turbulence parameter;} for different stopping times or particle size.  {  The dotted horizontal line in the second panel indicates the critical dust-to-gas ratio, above which the streaming instability is expected to operate efficiently.} 
The temperature profile and metallicity are fixed to $q=1$ and $Z=0.01$, respectively. 
\label{Hdust_varTstop}}
\end{figure}

Fig. \ref{Hdust_varTstop} compares the fiducial case with $\Tstop=10^{-3}$ to runs with $3$, $5$, and $10$ times that particle size. Increasing $\Tstop$ allows faster settling, since the settling speed $\left|v_{z,d}\right|\sim c_s \Tstop$ \citep{takeuchi02}. Larger particles can settle to a thinner layer before VSI-stirring, which begins at $t\sim 100P_0$ independently of $\Tstop$. That is, the initial VSI growth rate is insensitive to particle size.  
However, VSI-stirring weakens with increasing $\Tstop$, as 
$\avg{\Hdust/\Hgas}$ {  `bounce'} to smaller values with larger particles after the VSI activates. 
This is also reflected in the much smaller $\reyz$ with $\Tstop=0.005$ and $0.01$ compared to cases with $\Tstop=0.001$ and $0.003$. 


\begin{figure}
\includegraphics[scale=.22,clip=true,trim=0cm 4cm 0cm 0cm]{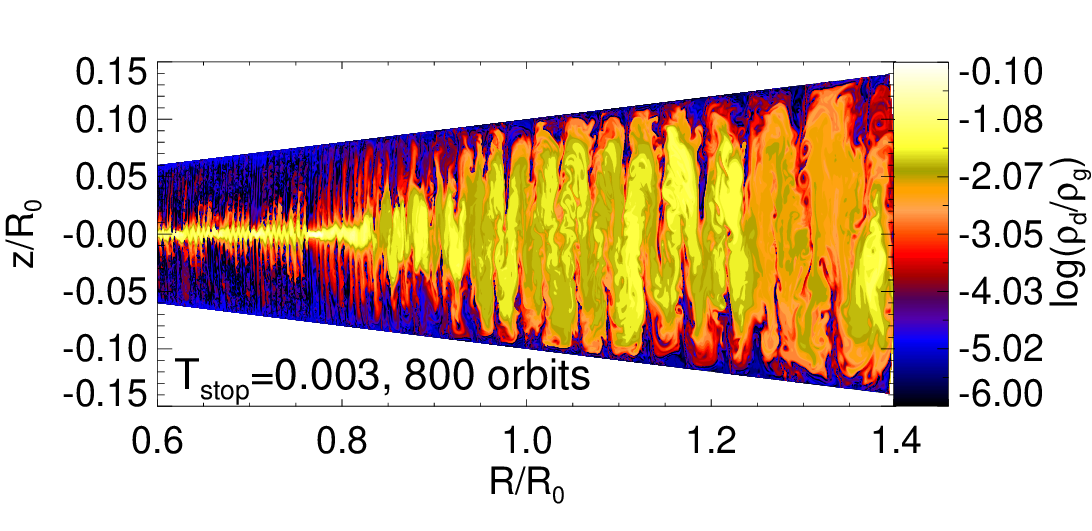}\\
\includegraphics[scale=.22,clip=true,trim=0cm 0cm 0cm 1.5cm]{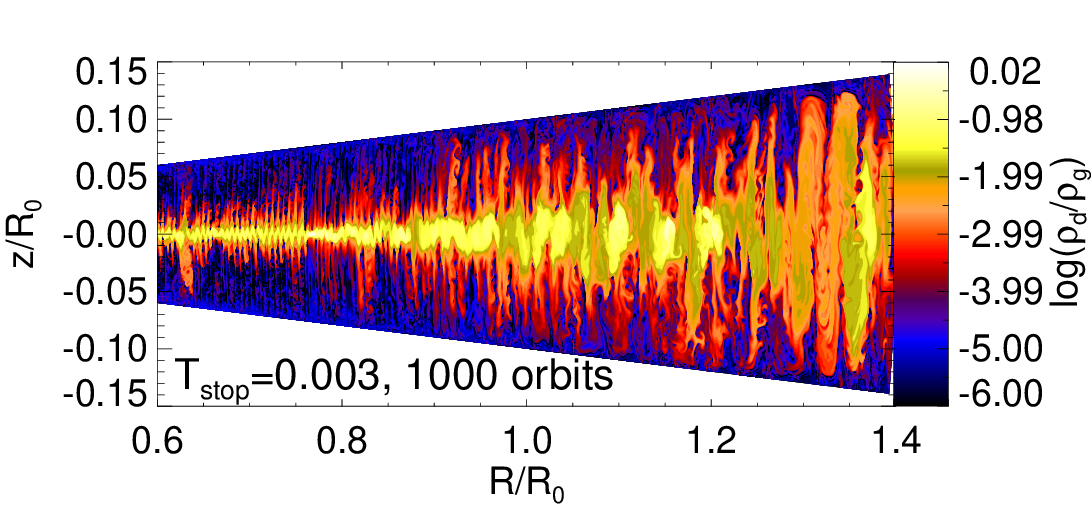}
\caption{Spontaneous settling in the run with $\Tstop=0.003$. The disc remains in quasi-steady state with vertically well-mixed dust until $t\sim 800P_0$, after which the dust begins to settle. 
\label{spont_settle}}
\end{figure}

In the $\Tstop=0.003$ run, dust is almost perfectly remixed by the initial VSI. A quasi-steady state is apparently attained, but only until $t\sim800P_0$, when we observe spontaneous settling. This is shown in Fig. \ref{spont_settle}, suggesting that the prior steady state was unstable. 
The $\Tstop=0.005$ and $\Tstop=0.01$ runs behave similarly. In both cases dust begins to settle soon after being remixed by the VSI. However, they also exhibit hints of spontaneous settling. The $\Tstop=0.005$ run attains a brief steady state ($150P_0\lesssim t\lesssim 250P_0$) before settling; while at the end of the $\Tstop=0.01$ run $\avg{\epsilon_0}$ suddenly increases. We discuss in \S\ref{feedback_settle} how spontaneous settling may be possible due to particle back-reaction. 

\begin{figure}
\includegraphics[scale=.22,clip=true,trim=0cm 4cm 0cm 0cm]{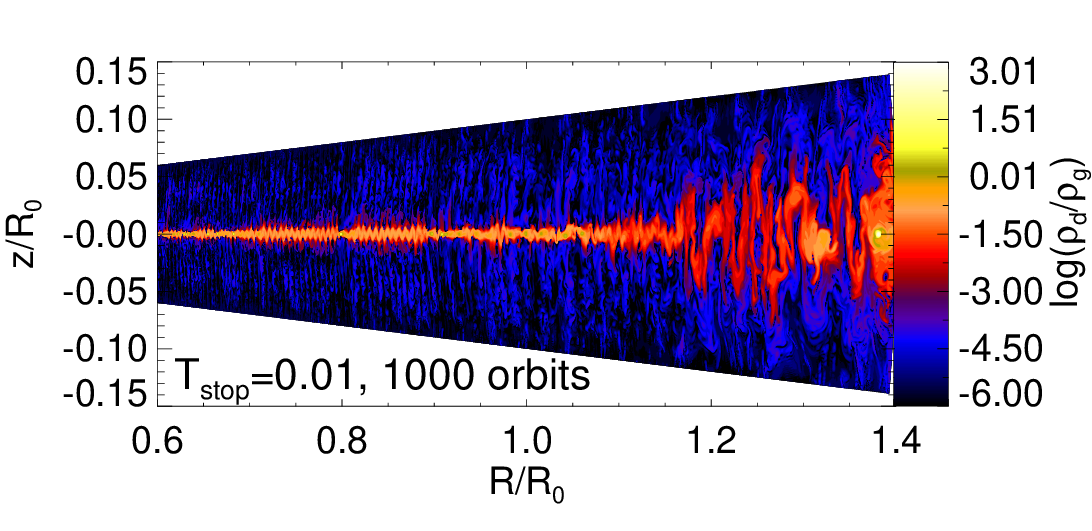}\\
\includegraphics[scale=.22,clip=true,trim=0cm 0cm 0cm 1.5cm]{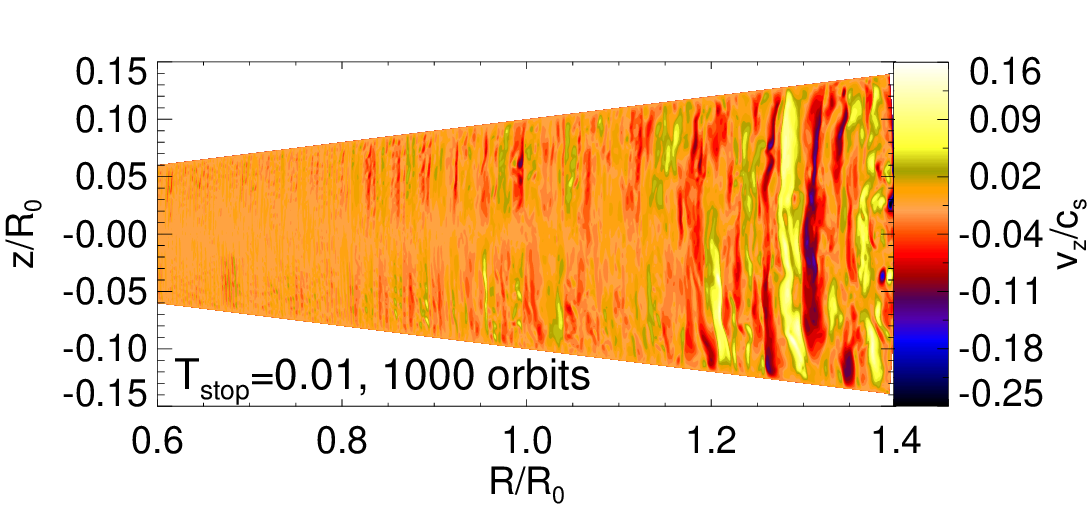}
\caption{Dust-to-gas ratio (top) and vertical velocity (bottom) at the end of the simulation with $\Tstop=10^{-2}$. 
\label{varTstop_dgratio_vz}}
\end{figure}

Fig. \ref{varTstop_dgratio_vz} shows the final dust-to-gas ratio and vertical velocity distribution in the $\Tstop=0.01$ run. A dust clump develops at the outer boundary, possibly due to boundary conditions, which may be related to the increased dust mixing in $R\gtrsim 1.1 R_0$. However, the sampling region around $R\simeq R_0$ is unaffected, and  particles there have settled to a thin layer with little vertical motion, due to self-stabilisation by dust-induced buoyancy \citepalias{lin17}.

\begin{figure}
\includegraphics[width=\linewidth]{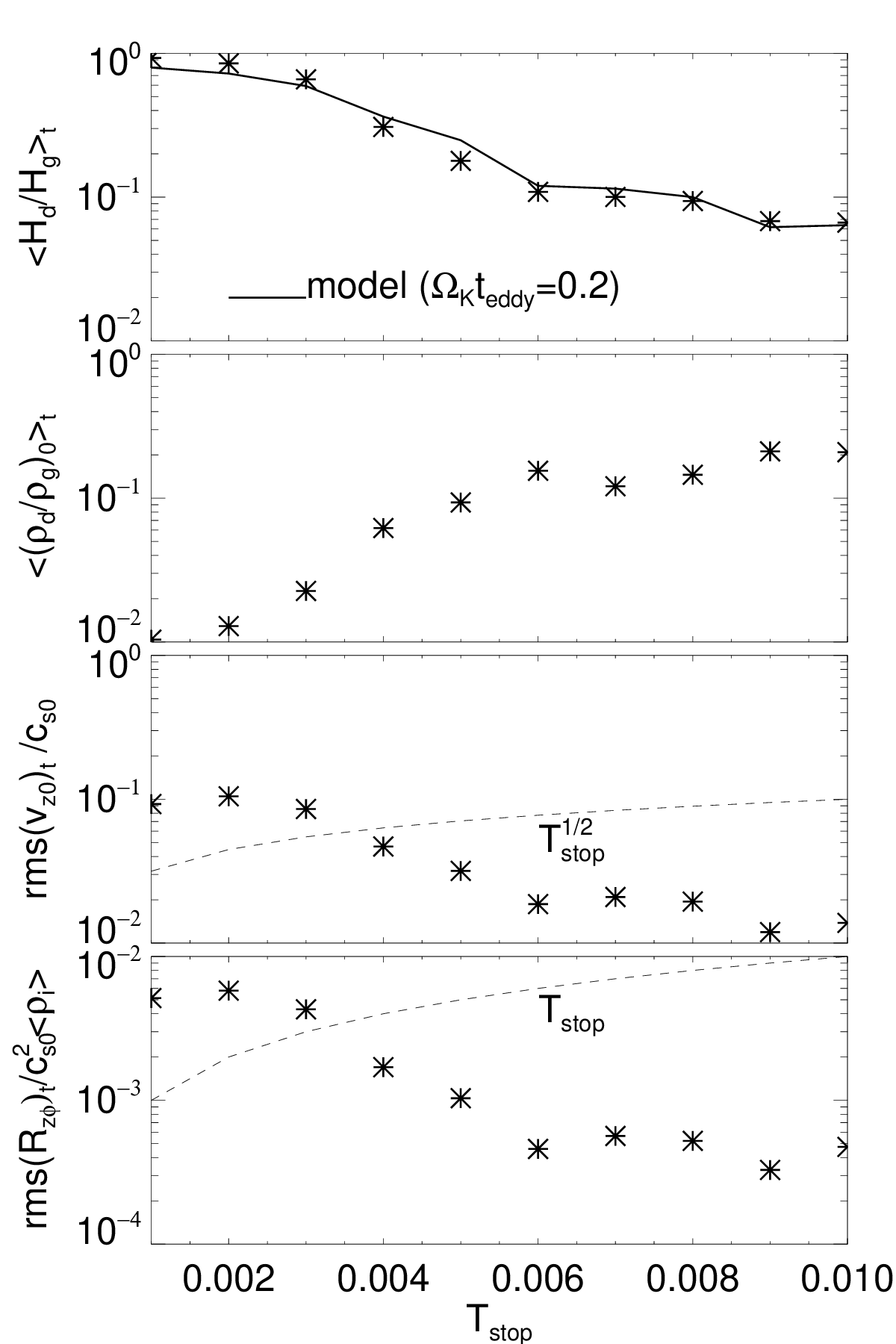}
\caption{ 
{  Top to bottom: time-averaged dust scale-height, midplane dust-to-gas ratio, midplane vertical velocity, and vertical turbulence strength;} as a function of the particle size parameterised by $\Tstop$. {  The solid line in the top panel is obtained from Eq. \ref{hdust_theory}.} The radial temperature gradient and metallicity is fixed to $q=1$, $Z=0.01$, respectively.  
\label{varTstop_Hdust_final}
}
\end{figure}

Fig. \ref{varTstop_Hdust_final} compares an extended set of simulations {  with} varying $\Tstop$. Here the time average is taken between the onset of the VSI and the end of the simulations. Significant settling occurs for $\Tstop \gtrsim 0.005$, for which $\avgt{\Hdust/\Hgas}$ drops to $\sim 0.1$ and $\avgt{\epsilon_0}\sim O(0.1)$. 
{ 
 The measured $\avgt{\Hdust/\Hgas}$ is consistent with the model described by Eq. \ref{hdust_theory} if we choose $\OmK\teddy=0.2$, again similar to eddy timescales found by \cite{stoll16}.   
}
Although less clear-cut, this figure {  also indicates that dust settling occurs when $\sqrt{\avg{\vzmid^2}}_t\lesssim \sqrt{\Tstop} c_{s0}$ or $\reyzt \lesssim \Tstop$. 
}


\subsection{Effect of dust abundance}


{  Dust-induced buoyancy is expected to help} settling by stabilising the VSI {  (\citetalias{lin17})}.  Here we investigate this effect more precisely by varying the solid abundance or metallicity $Z=\Sigd/\Sigg$. Larger $Z$ corresponds to stronger {  dust} feedback {  onto the gas}. For our fiducial parameter values $(q,\Tstop,Z)=(1,10^{-3},10^{-2})$ dust does not settle. We thus consider $Z\in[0.01,0.1]$ in this section. In Appendix \ref{nofeedback} we {  demonstrate 
the VSI is particularly sensitive to dust-loading in weakly turbulent discs, in which case the stabilising  effect of feedback manifests even with $\epsilon_0$ of a few per cent. } 

\begin{figure}
\includegraphics[width=\linewidth]{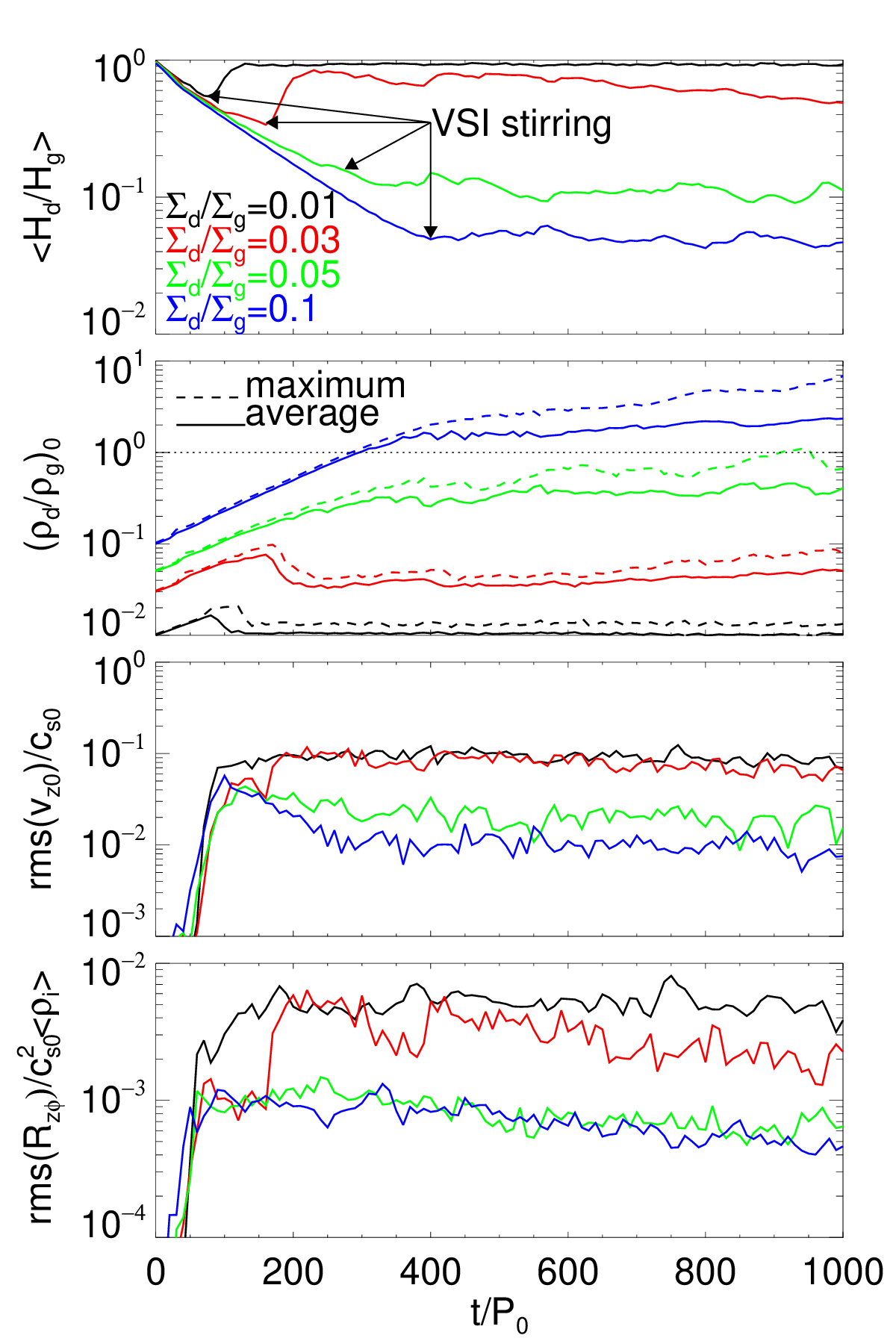}
\caption{ 
{  Top to bottom: evolution of dust-scale height, midplane dust-to-gas ratios, midplane vertical velocity, and vertical turbulence parameter;} for different metallicities. {  The dotted horizontal line in the second panel indicates the critical dust-to-gas ratio, above which the streaming instability is expected to operate efficiently.} The radial temperature gradient and particle size is fixed to $q=1$ and $\Tstop=10^{-3}$, respectively. \label{Hdust_varZ}}
\end{figure}

Fig. \ref{Hdust_varZ} compares the disc evolution for $Z=0.01$, $0.03$,  $0.05$, and $0.1$. Increasing $Z$ does not affect the settling timescale. Instead, larger $Z$ delays the VSI, which allows the dust to settle to a thinner layer. This reflects the stabilising effect of particle  back-reaction. For $Z=0.03$, the VSI remixes the dust at $t\sim 200P_0$, and settles slowly afterwards. For $Z=0.05$ and $Z=0.1$, the initial VSI only halts the settling process, and does not increase $\avg{\Hdust/\Hgas}$, unlike the $Z=0.01$ and $Z=0.03$ runs.    

For $Z=0.1$, we find $\avg{\epsilon_0} \sim  2$ for $t\gtrsim 300P_0$. This is expected to develop the SI, but our numerical {  setup and} resolution is insufficient to resolve the SI for the small particles considered here \citep{yang16}. In the $Z=0.05$ and $Z=0.03$ runs,  $\avg{\epsilon_0}$ does not reach unity within the simulation timescale, but is still significantly enhanced relative to the $Z=0.01$ run. Comparing these runs to 
the previous simulations with increased $\Tstop$ (Fig. \ref{Hdust_varTstop}) show that 
increasing $Z$ is more effective in raising {  the dust-to-gas ratio} than increasing particle size. This is not surprising since, for a given $\Hdust$ a larger $Z$ implies a larger $\epsilon_0$ {  (Eq. \ref{Z_def})}; let alone in the present cases where a smaller {  dust scale-height} is reached with increasing $Z$ compared to the same increase in $\Tstop$. 


The $Z=0.05$ and $Z=0.1$ cases show {  much reduced midplane vertical velocities and} vertical turbulence compared to $Z=0.01$, which is consistent with the former cases reaching much smaller values of $\avg{\Hdust/\Hgas}$. The $Z=0.1$ run also displays a slow drop in $\reyz$ for $t\gtrsim400P_0$, with a corresponding rise in $\avg{\epsilon_0}$; while $\avg{\Hdust/\Hgas}$ remains relatively constant. This signifies particle settling into midplane unhindered; in fact this settling is stabilising the system. 
The $Z=0.03$ run attains $\reyz\simeq 5\times10^{-3}$ early on, similar to the $Z=0.01$ case, but it eventually drops to $\reyz\simeq 2\times10^{-3}$. This also reflects the slow settling observed after the initial VSI in that case. 



\begin{figure}
\includegraphics[scale=.22,clip=true,trim=0cm 4cm 0cm 0cm]{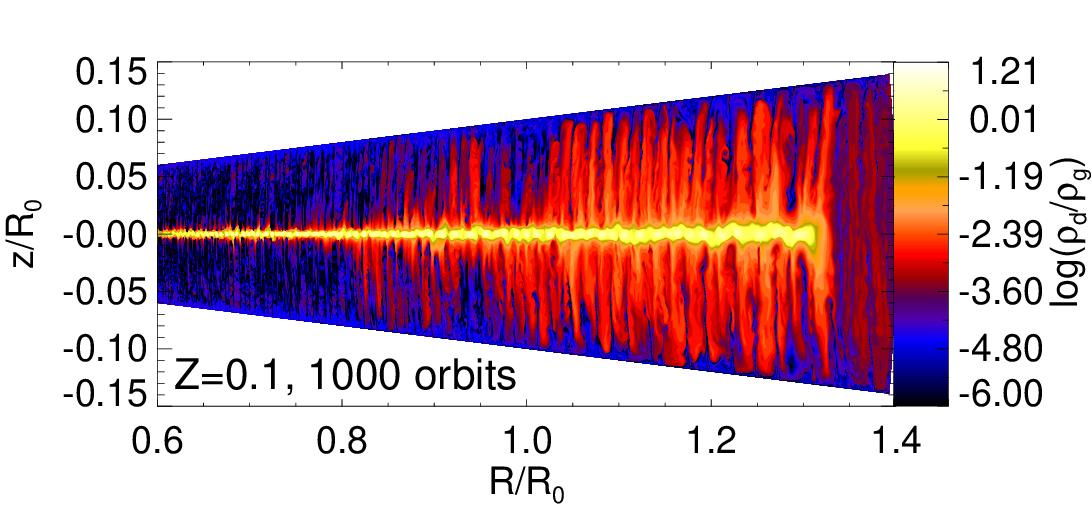}\\
\includegraphics[scale=.22,clip=true,trim=0cm 0cm 0cm 1.5cm]{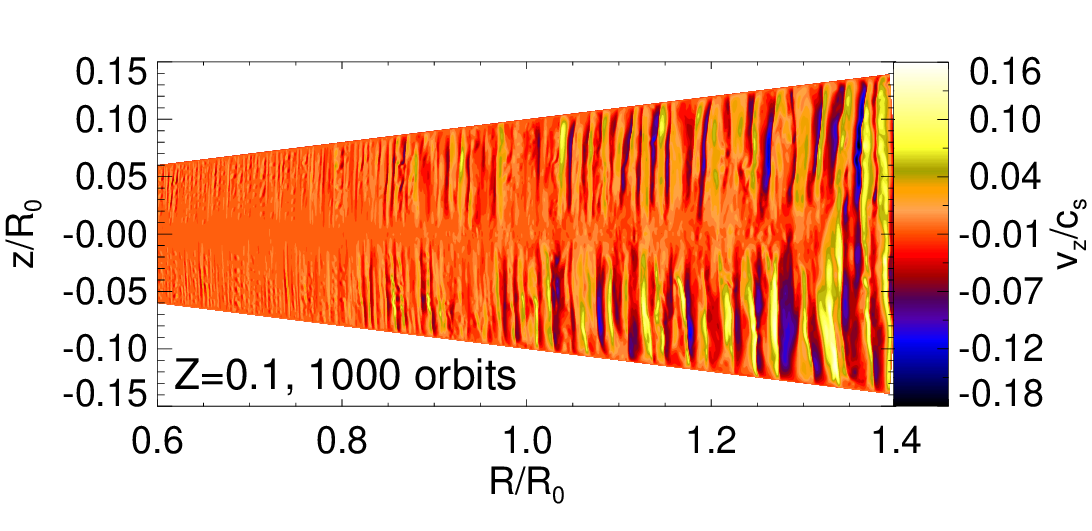}
\caption{Dust-to-gas ratio (top) and vertical velocity (bottom) at the end of the simulation with $Z=0.1$. 
\label{dgratio_q1_Z0d1}}
\end{figure}

Fig. \ref{dgratio_q1_Z0d1} shows $\rhod/\rhog$ and $v_z$ at the end of the $Z=0.1$ simulation. {  Maximum vertical} velocities are reduced by at a factor of two from the standard case with $Z=0.01$ (Fig. \ref{q1_final_vz}), {  while midplane values are reduced by an order of magnitude} (Fig. \ref{Hdust_varZ}). Similar to the large-particle simulation (Fig. \ref{varTstop_dgratio_vz}), the dust  midplane is quiescent, although not completely inactive. This dense dust layer also decouples the upper, dust-free region from that in the lower disc. 

\begin{figure}
\includegraphics[width=\linewidth]{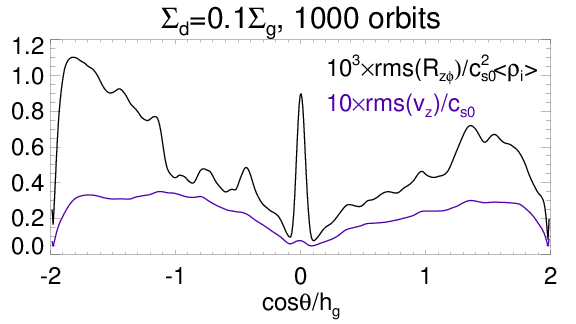}
\caption{Vertical stress {  (black) and vertical velocity (purple) profiles} at the end of the $Z=0.1$ simulation.  
\label{alpha_z_Z0d1}}
\end{figure}

Fig. \ref{alpha_z_Z0d1} shows the corresponding vertical stress {  and vertical velocity profiles}. 
Vertical stresses generally increase away from the midplane, reaching $\reyz\sim 10^{-3}$ near the boundaries{ . This reflects VSI activity in the dust-poor layers above and below the midplane.} 
The spike at $z=0$ is associated with {  the fact that the dust layer rotates} closer to the Keplerian speed compared to the initially gas-dominated state. {  (Recall from Eq. \ref{alpha_z_def} that $R_{z\phi}\propto \Delta v_\phi$.)} {  Although vertical velocities are reduced towards the midplane, notice that $\vzmid\neq 0$. Such residual midplane vertical motions are 
 } 
  due to disturbance by bulk vertical motions from the VSI-active layers {  (the midplane dust layer in itself is expected to be stable, see below)}. This explains why the dust layer maintains a finite thickness and does not settle further. 

\subsubsection{Dust-induced buoyancy}

\begin{figure}
\includegraphics[width=\linewidth]{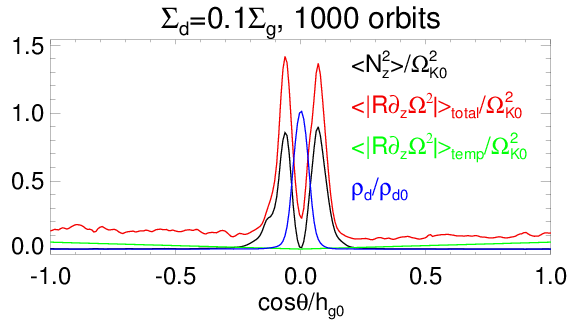}
\caption{Vertical profiles in the squared vertical buoyancy frequency (black, Eq. \ref{dusty_buoyancy}), vertical shear rate (total: red; temperature-related: green,  Eq.\ref{vshear_Tgrad}), and dust density (blue {,  normalised to the midplane value}) at the end of the $Z=0.1$ simulation. 
\label{vshear_Z0d1}}
\end{figure}

Fig. \ref{vshear_Z0d1} compares the vertical buoyancy frequency and vertical shear profiles at the end of the $Z=0.1$ simulation. We show both the actual vertical shear rate, $\left|R\p_z\Omega^2\right|$; and that associated with the imposed temperature gradient{  
, $\left|R\p_z\Omega^2\right|_\mathrm{temp}$, as estimated from Eq. \ref{vshear_Tgrad}}. 
This quantity is the driver of the underlying VSI \citepalias{lin17} and is essentially constant in time. 
{  On the other hand, the total} vertical shear displayed in Fig. \ref{vshear_Z0d1} is {  either dominated by} the settled dust in the midplane, or {  by} VSI modes away from it. 

{  The total vertical shear exceeds buoyancy throughout the disc column.} 
However, recall in axisymmetric discs {  that} dust-induced vertical shear \emph{does not} cause instabilities \citepalias{lin17}. Thus in our simulations instability is associated with vertical shear induced by the imposed temperature gradient, {$\left|R\p_z\Omega^2\right|_\mathrm{temp}$}. {  Away from the midplane at $|z|\gtrsim0.1\Hgas$, buoyancy frequencies vanish, whereas the temperature-related, destabilising vertical shear rate is larger there. About the midplane ($|z|\sim 0.1\Hgas$), however, this destabilising effect is overwhelmed by the stabilising buoyancy.} This explains the `layered' structure seen in Fig. \ref{dgratio_q1_Z0d1}: a quiet midplane with VSI-active layers above and below it.   



\begin{figure}
\includegraphics[width=\linewidth]{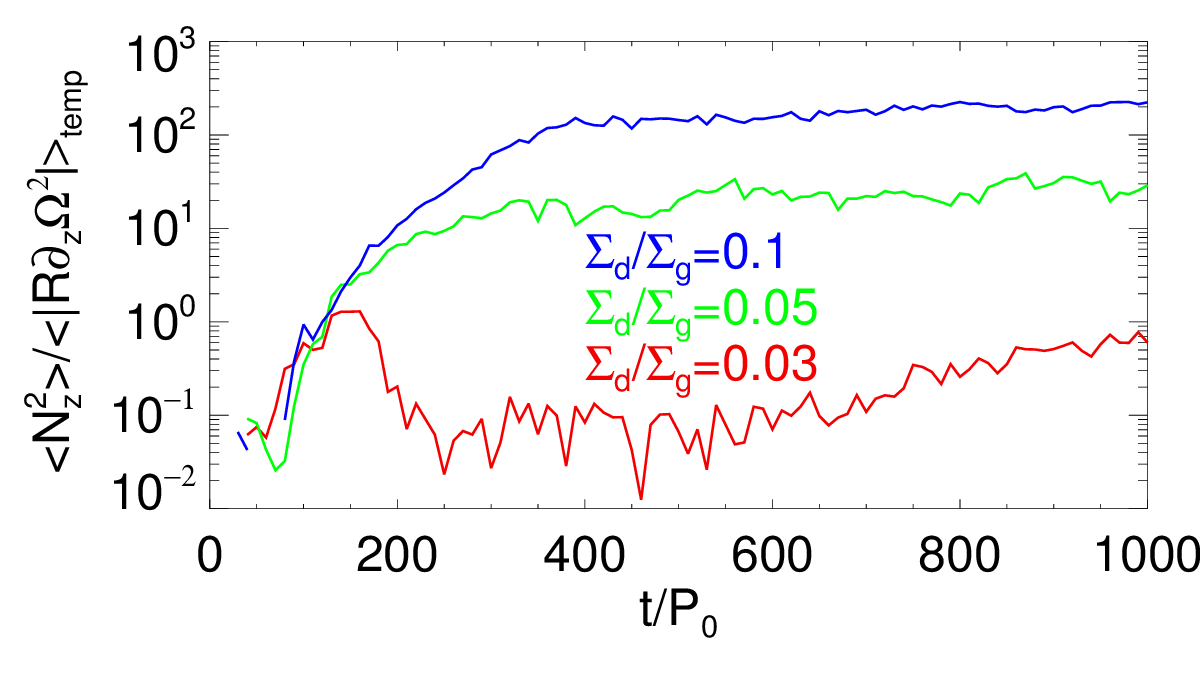}
\caption{Comparison between the stabilising effect of dust-induced buoyancy (Eq. \ref{dusty_buoyancy}) to the destabilising effect of the temperature-related vertical shear (Eq. \ref{vshear_Tgrad}), as a function of the metallicity. The averages are taken over the dust layer. 
\label{vshear_compareZ}}
\end{figure}

 Fig. \ref{vshear_compareZ} compares the stabilising buoyancy to the destabilising temperature-related vertical shear, averaged over the dust layer, for several cases. We omit the $Z=0.01$ run here since $N_z^2\sim 0$ in that case. For $Z=0.03$ this ratio is $\lesssim 1$, though rising later on, which is consistent with the slow settling. For $Z=0.05$ and $Z=0.1$ this ratio is $\gg 1$, and the VSI is strongly stabilised by buoyancy within the dust layer.

\subsubsection{Dust settling by dust-loading}

\begin{figure}
\includegraphics[width=\linewidth]{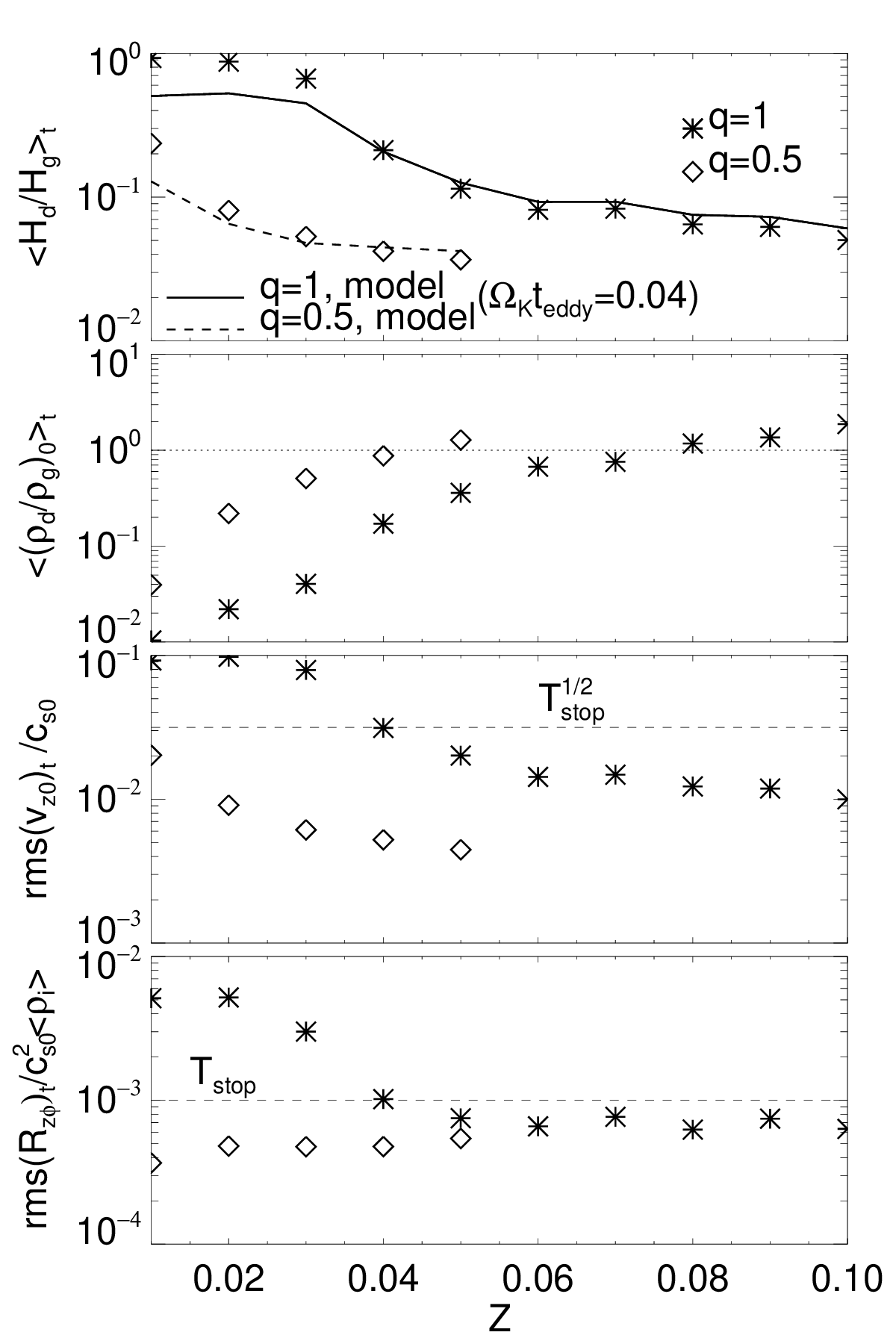}
\caption{
{  Top to bottom: time-averaged dust scale-height, midplane dust-to-gas ratio, midplane vertical velocity, and vertical turbulence strength;} 
as a function of the metallicity. The radial temperature gradient is $q=1$ (asterisks) and $q=0.5$ (diamonds). {  The solid (dashed) line in the top panel is obtained from Eq. \ref{hdust_theory} for the $q=1$ ($q=0.5$) cases. The dotted horizontal line in the second panel indicates the critical dust-to-gas ratio, above which the streaming instability is expected to operate efficiently.}
The particle size is fixed to $\Tstop=10^{-3}$.  
\label{varZ_Hdust_final}
}
\end{figure}

We now examine the critical metallicity for dust settling to overcome the VSI. Fig. \ref{varZ_Hdust_final} compare {  quasi-steady states} as a function of $Z$. We also plot several cases with $q=0.5$. 

For $q=1$ we find stabilisation by dust-loading is most dramatic for { $Z\simeq 0.03$---$0.05$; while for $Z > 0.05$ dust settles to $\avgt{\Hdust/\Hgas}\lesssim 0.1$ with little variation with respect to $Z$}. On the other hand, for $q=0.5$ dust can settle to $\avgt{\Hdust/\Hgas}\lesssim 0.1$ with $Z\gtrsim 0.02$; suggesting the critical $Z$ for settling scales as $\sim 1/|q|$. {  We find the settled dust scale-heights can be fitted with Eq. \ref{hdust_theory}, but this required $\OmK\teddy \simeq 0.04$, which is smaller than that found by \cite{stoll16}. This indicates that increasing the metallicity reduces the turbulence eddy timescales. We find  $\sqrt{\avg{\vzmid^2}}_t\sim \sqrt{\Tstop}c_{s0}$ continues to distinguish between settled and well-mixed cases as we vary the metallicity. 
}

{  In the settled cases, we find} the dust layer is too thin to affect {  the average vertical stress} $\reyzt$, i.e. the measured vertical turbulence is associated with the gas layers above and below the midplane, hence $\reyzt$ flattens off. {  Nevertheless, we find settled cases still obey $\reyzt \lesssim \Tstop$.}




These results indicate that in weakly VSI-turbulent discs, dust settling is sensitive to the local metallicity. This implies that small changes in $Z$, for example {  due to} radial drift {  and pile-up} of dust, could have significant impact on dust settling in disc regions with shallow temperature profiles (see \S\ref{discuss} for a discussion). 


\section{Discussion}\label{discuss}


\subsection{Overcoming VSI turbulence through particle back-reaction}

Our simulations indicate that particle back-reaction onto the gas is important for dust settling in discs prone to the VSI.  {  This implies that previous studies using passive particles may have overestimated the minimum particle sizes that can settle in VSI-active discs \citep{stoll16,flock17}.} 





We find that raising the metallicity $Z=\Sigd/\Sigg$, which increases the particle back-reaction, helps dust settling. In this case the settling speed is unaffected.  Instead, VSI stirring is increasingly delayed due to stabilisation by dust-induced buoyancy forces. This results in thinner dust layers before the VSI halts the settling process, which are even more difficult to stir up. (The same effect occurs for increasing particle size.) 

For $\Tstop=10^{-3}$ and $q=1$ we find $Z\gtrsim 0.05$ is needed for dust to settle down to $\Hdust\lesssim 0.1\Hgas$; while in less turbulent discs with $q=0.5$ this only required $Z\gtrsim 0.02$, slightly above the solar value. This suggests that in weakly turbulent discs the VSI is sensitive to the disc metallicity. {  This is consistent with the argument presented in \S\ref{buoyant_stabilisation} and the simulation in Appendix \ref{nofeedback}, which show} that the VSI can be stabilised when the dust-to-gas ratio $\epsilon$ {  becomes comparable to the disc's aspect-ratio $\hgas$, i.e. } a few per cent {  in PPDs.} 

Our simulations show that $\Hdust$ generally depends on $q$, $\Tstop$, and $Z$. All three parameters affect the VSI turbulence strength {  $\reyz$ (the dimensionless vertical Reynolds stress) and the midplane vertical velocity $\vzmid$. We find that the settled and well-mixed cases are separated by $\reyz \sim \Tstop$ or $\sqrt{\avg{\vzmid^2}}\sim \sqrt{\Tstop}c_{s}$. The latter condition is consistent with the settling model of \citet{dubrulle95}, provided an appropriate eddy timescale is chosen (see Eq. \ref{hdust_theory}). 

We can apply the above results to assess whether particles of a given $\Tstop$ ($>0$) and abundance $Z$ can settle.  Recent analyses indicate $\left| v_z\right| \sim \hgas |q| c_s$ for VSI turbulence \citep{barker15, latter18}, which is consistent with our simulations. We thus expect particles with $\Tstop \gtrsim \hgas^2q^2$ to settle, even in the limit $Z \to 0$. However, if $\Tstop \lesssim \hgas^2q^2$, then $Z \gtrsim  \hgas|q|$ is needed to weaken the VSI and enable dust settling (see \S\ref{buoyant_stabilisation}).   
}


\subsection{Self-sustained settling}\label{feedback_settle}

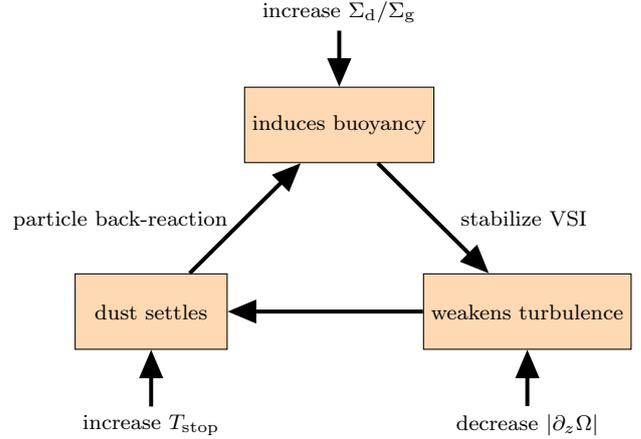
\begin{figure}	
\tikzstyle{process} = [rectangle, minimum width=2cm, minimum height=1cm, text centered, draw=black, fill=orange!30]
\tikzstyle{effect} = [rectangle, minimum width=2cm, minimum
height=0.2cm, text centered, draw=none, fill=none]
  \tikzstyle{line} = [draw, -triangle 45,thick,line width=1.5]
  \begin{center}
    \begin{tikzpicture}[node distance=5cm]
      \node[process](settle){dust settles};
      \node[process,above right of=settle, node distance = 3.5cm](nz){induces buoyancy};
      \node[process,below right of=nz, node distance = 3.5cm](vsi){weakens turbulence};
       \node[effect,below of =settle, node distance =
      1.5cm](psize){increase $\Tstop$};
       \node[effect,below of=vsi, node distance =
      1.5cm](varq){decrease $\left|\p_z\Omega\right|$};
       \node[effect,above of=nz, node distance =
      1.5cm](metal){increase $\Sigd/\Sigg$};
      \path [line](settle) -- node [xshift=-5.5em,yshift=-.1em]
      {particle back-reaction} (nz);
      \path [line](nz) -- node [xshift=4em]
      {stabilize VSI} (vsi);
      \path [line](vsi) -- (settle);
      \path [line](psize) --node[]{}(settle);
      \path [line](metal) --node[]{}(nz);
      \path [line](varq) --node[]{}(vsi);
    \end{tikzpicture}
 \end{center}
\caption{Self-sustained settling of dust in VSI-active discs. A positive feedback loop 
is enabled by particle back-reaction onto the gas. This allows dust settling to induce buoyancy forces to stabilise the VSI, which reduces the turbulent stirring, leading to further settling. 
\label{settle_feedback}}
\end{figure}

We propose particle back-reaction can lead to self-sustained dust settling, as illustrated in  Fig. \ref{settle_feedback}. 
In this picture, dust settles if it can induce an effective buoyancy that stabilises the VSI, which favours further settling, and the cycle repeats.   
This feedback cycle may be initiated by increasing $\Tstop$ via particle growth, raising  the local metallicity $Z$ via radial dust drift, or reducing the radial temperature gradient $|q|$. The last trigger is less likely in PPDs at large radii because the local temperature profile is fixed by stellar irradiation. 

We suggest this positive feedback loop is responsible for spontaneous settling observed in some simulations (e.g. Fig. \ref{spont_settle}). The quasi-steady balance between VSI-stirring and dust settling is delicate because the VSI is sensitive to buoyancy. If the system is perturbed such that $Z$ increases locally, then that region may begin to settle and enter the feedback loop. 

However, self-sustained settling does not continue indefinitely. As dust settles to the midplane, it leaves behind dust-free, VSI-turbulent layers above and below it. Vertical motion from the active layers continue to disturb/distort the midplane dust layer, eventually halting further sedimentation. 

Indeed, we find that although $\left|v_z\right|$ drops significantly with decreasing $|z|$, it is still non-zero about the dusty midplane. Such a layered structure {  in $\left|v_z\right|$} is characteristic of settled dust in VSI-prone discs. {  In fact, a similarly layered structure is found in the linear VSI with a dusty midplane (\citetalias{lin17}, see their section 6.3). 
This picture is analogous to layered accretion discussed by \cite{gammie96}; although in that case it is the radial mass accretion rate that is minimised in the magnetic dead zones near the midplane.} Dust settling in the dead zone is halted by {  residual activity induced by the turbulent active layers} \citep{fromang06}. 
{  We suggest a similar effect is present in our simulations: the VSI active layers induce residual vertical motions at the disc midplane to maintain a finite dust layer thickness.}

In reality, self-sustained settling may also be terminated by other processes, such as the KHI \citep{chiang08,lee10}, GI \citep{goldreich73}, or SI \citep{youdin05a}. However, these effects are not captured by our numerical models.  


\subsection{Implications for planetesimal formation} 


Efficient planetesimal formation via rapid SIs require a midplane dust-to-gas ratio of order unity or larger \citep{youdin05a,johansen09}.   

However, our results indicate small particles would remain lofted by VSI turbulence. Here, what constitutes as `small' depends on the turbulence strength and local metallicity. 
In the outer parts of PPDs, VSI turbulence is primarily related to the local temperature gradient $|q|$. If we take $|q|\simeq 1$ and solar metallicity $Z=0.01$, then by small we mean particles with $\Tstop$ of $O(10^{-3})$. At a distance of $20$au, this roughly corresponds to particle sizes of $O(10^2\mu\rm{m})$, assuming an internal density $\rho_\bullet = 1\,\rm{g}\rm{cm}^{-3}$ (see Eq. \ref{tstop_est}). 

In the above example, particles smaller than $\sim 100\mu\rm{m}$ will remain lofted. However, if they can grow to $\sim $mm sizes amidst the turbulence, {  e.g. by sticking} \citep{blum18}, or if the local metallicity increases \citep[e.g. due to a radial pile-up of dust, ][]{kanagawa17}; then these particles can settle against the VSI and eventually undergo SI. 

Raising the local metallicity is an effective way to achieve {  a local dust-to-gas ratio} $\rhod/\rhog\sim 1$ in two regards. First, a larger $Z$ allows $\epsilon_0\sim 1$ to be attained at larger $\Hdust$ (see Eq. \ref{Z_def}). That is, the required degree of settling is smaller. Second,  increasing $Z$ actively stabilises the disc against the VSI, thereby allow for effective settling.  

Coincidentally, non-linear numerical simulations of the SI indicate $Z\gtrsim 0.02$ is required for strong clumping {  of particles with $\Tstop\sim 0.01$---$0.1$} \citep{johansen09,bai10}{ ; while smaller particles with $\Tstop\sim 10^{-3}$ require $Z\gtrsim 0.04$ \citep{yang16b}.} {  These particle clumps} can then undergo GI. One also requires $Z\gtrsim 0.03$ in order to overcome the KHI associated with dust-induced vertical shear, which would overturn the dust layer \citep{chiang08,lee10}.   


We conclude that in the outer parts of protoplanetary discs, small particles must have super-solar metallicities in order to form planetesimals self-consistently{ : dust can then settle against VSI turbulence, trigger efficient streaming instability, clump, and finally undergo gravitational collapse.} 

\subsection{Application to observations}




Our simulations may aid the interpretation of PPD observations. As a concrete example, we consider the PPD around HL Tau \citep{alma15}. This inclined disc displays well-defined dust rings at all azimuth, which implies that the dust has settled. 

For the HL Tau disc, \cite{pinte16} estimates $\Hdust\sim 0.07\Hgas$ at $100$au, and   $\epsilon_0\sim 0.2$ assuming $Z=0.01$. They infer a 
{  dimensionless vertical diffusion parameter} $\alpha \simeq 3\times 10^{-4}$, assuming passive particles  \citep{fromang09}. 
Interestingly, these values are consistent with our $\Tstop=0.01$ simulation\footnote{Note that we used $(q,\hgas)=(1,0.05)$ while \citeauthor{pinte16} and \citeauthor{jin16} assumed $(q,h_\mathrm{g})=(0.7,0.1)$ and $(0.43,0.096)$, respectively. However, the destabilising effect of increasing $|q|$ offsets the stabilising effect of decreasing $\hgas$ \citep{nelson13}.} 
shown in Fig. \ref{varTstop_Hdust_final}, {  if we identify $\alpha$ as our dimensionless vertical Reynolds stress.} 
Using Eq. \ref{tstop_est}, $\Tstop=0.01$ corresponds to particle sizes $\sim90\mu$m at $100$au, similar to that adopted in the HL Tau disc model of \cite{jin16}. For mm-sized  particles, $\Tstop=0.01$ requires a disc $\sim 10$ times more massive than the MMSN; or grains with internal density $\sim 0.1\mathrm{gcm}^{-3}$.  


For dusty VSI turbulence we find $\Hdust=\Hdust(q, \Tstop, Z)$. This may be used to estimate the local metallicity. For example, let us assume $q \simeq 0.5$ and particle sizes of $\sim 100\mu\rm{m}$, and a massive (young) disc with $F=10$. This gives $\Tstop\simeq 10^{-3}$ at $100$au. Then Fig. \ref{varZ_Hdust_final} suggest \citeauthor{pinte16}'s estimate, $\Hdust\simeq 0.07\Hgas$, can be achieved with $Z\simeq 0.03$. {  On the other hand, if $Z\sim 10^{-3}$ so that particle back-reaction is negligible, then VSI turbulence would be too strong to allow dust to settle \citep[see, e.g.][]{flock17}.}



The interplay between VSI turbulence, dust settling, and the local metallicity may lead to a radially-dependent dust scale-height, as follows. Suppose dust drifts radially, say from $R_2$ to $R_1$ such that $Z(R_1)>Z(R_2)$. { 
Then the increased (decreased) metallicity at $R_1$ ($R_2$) would stabilise (destabilise) the disc there. 
Now, if the gas density contrast between $R_1$ and $R_2$ is smaller than that in $Z$, i.e. $Z_1/Z_2 > \Sigma_\mathrm{g1}/\Sigma_\mathrm{g2}$, so that the contrast in stopping times is outweighed by that in metallicity, then dust settling (mixing) should be promoted at $R_1$ ($R_2$).}
Ultimately, $\Hdust(R_1)<\Hdust(R_2)${ , assuming similar gas scale heights if $R_1>R_2$.} That is, dust rings should have {  smaller} scale-heights compared to dust gaps\footnote{  However, if the gas density in the dust ring is also increased significantly, then the decreased stopping times may negate the positive effect of increased metallicity on settling.}. This can be tested with future observations of the radial and vertical dust structure in PPDs.

\subsection{Caveats and outlooks}
Below we discuss several caveats of our modelling and directions for future work. 

We assumed axisymmetry throughout. This simplification overestimates the strength of VSI turbulence \citep{manger18}. 
For example, \cite{stoll16} found that passive particles with $\Tstop=0.002$ can settle to $\Hdust\simeq 0.66\Hgas$; whereas we find such particles remain perfectly mixed with the gas. On the other hand, \citeauthor{stoll16} found particles with $\Tstop=0.02$ only settles to $\Hdust\simeq0.4\Hgas$, whereas we find $\Tstop=0.01$ particles can already settle to $\Hdust\lesssim 0.1\Hgas$, which could be due to the stabilising effect of particle back-reaction in our case. 

Axisymmetric models also suppress vortex formation due to the VSI \citep{richard16,manger18}, as well as the KHI due to the vertical shear between the dust and gas layers \citep{chiang08}. Vortices are known to act as effective dust traps \citep{lyra13}, although they can also source hydrodynamic turbulence \citep{lesur09b,lesur10}.  The KHI may become relevant if the dust layer becomes sufficiently thin and/or if the overall metallicity is large (see Appendix \ref{KHI}). 
It will be necessary to perform full 3D simulations to investigate how these additional processes affect dust settling. 


We adopted initial disc models that minimised radial disc evolution in order to focus on vertical dust settling (\S\ref{model}, see also \citetalias{chen18}). However, in PPDs dust typically drifts inward. This may be important for dust settling because radial pile-up of dust, should it occur, would stabilise the VSI and induce dust settling. Future work should explore different initial dust and gas density profiles. 

The dust-gas model of \citetalias{lin17} only allows for a single species of dust particles. In reality, there will be a distribution of particle sizes, each with a different metallicity. Particles that are larger or more abundant will settle first, which is expected to stabilise the system. The settling of smaller particles may then ensue. To investigate this scenario, one would need a dust-gas framework that allows for multiples species of dust grains. 



\section{Summary}\label{summary}

In this paper we study the sedimentation of small dust particles in the outer parts of protoplanetary discs subject to hydrodynamic turbulence driven by the vertical shear instability {  (VSI)}. 
We perform 2D, axisymmetric numerical simulations to investigate the effect of {  VSI} turbulence strength, $\reyz$ {  (essentially the vertical Reynolds stress-to-pressure ratio)}; particle size or {  Stokes number}, $\Tstop$; and solid abundance (metallicity) $\Sigd/\Sigg$ on dust settling. 

We model VSI turbulence explicitly by imposing a fixed radial temperature gradient $q$. However, we find {  the resulting stress} $\reyz$ can also depend on $\Tstop$ and $\Sigd/\Sigg$. Nevertheless, our results indicate {  that dust settles when} the resulting $\reyz \lesssim \Tstop$ { . In our simulations this is equivalent to having {  a midplane vertical velocity dispersion} $\sqrt{\avg{\vzmid^2}}\lesssim \sqrt{\Tstop} c_s$. The latter {  condition} is consistent with the classic dust settling model of \cite{dubrulle95}.} 



We find that dust settling is favoured by 
\begin{inparaenum}[1)]
\item weaker VSI turbulence, corresponding to shallower radial temperature profiles;
\item larger particles; and 
\item increased local metallicity. 
\end{inparaenum}
{  We expect dust can settle if $\Tstop \gtrsim \hgas^2q^2$ and/or $\Sigd \gtrsim \hgas|q|\Sigg$, where $\hgas$ is the disc aspect-ratio.} 

Our simulations show that particle back-reaction plays an important role for dust settling against the VSI. This is because dust-loading, specifically vertical gradients in the dust-to-gas ratio, introduces a buoyancy force \citepalias{lin17}, which is known to be effective in stabilising the VSI \citepalias{lin15}. {  Thus settled dust is strongly self-stabilised against the VSI.} This results in a quiescent dusty midplane with residual {  vertical velocities induced by} VSI turbulence {  from} the gaseous layers above and below it, which maintains the dust layer thickness. 


We find dust settling {  can overcome the VSI by increasing the local} metallicity. For particles with $\Tstop=10^{-3}$, corresponding to grain sizes of $\sim 100\mu$m at $20$au in MMSN-like discs, we find that $\Sigd\gtrsim  0.04\Sigg$ enables dust reach $\rhod\sim\rhog$ at the disc midplane. Such metallicities are also expected to trigger planetesimal formation via the streaming instability \citep{johansen09,yang16b}. 
Our results are thus consistent with the notion that 
planetesimals can form from small dust particles in the outer parts of protoplanetary discs, provided that the corresponding solid abundance is a few times larger than the solar value. 

\section*{Acknowledgements}
{  I thank the anonymous referee for key suggestions that lead to several improvements to this paper. I am indebted to Eugene Chiang for valuable comments on an early draft of this work, especially for motivating \S\ref{buoyant_stabilisation}. I also thank Guillaume Laibe for useful discussions.  
} 
This work is supported by the Theoretical Institute for Advanced
Research in Astrophysics (TIARA) based in the Academia Sinica
Institute for Astronomy and Astrophysics (ASIAA) and 
the Ministry of Science and Education (MOST) grant   107-2112-M-001-043-MY3.   
Simulations were carried 
out on the TIARA High Performance Computing cluster and the TAIWANIA cluster hosted by 
the National Center for High-performance Computing (NCHC). I am grateful to the NCHC for computer time and facilities.

\bibliographystyle{mnras}
\bibliography{ref}

\appendix
\section{Dust settling with negligible back-reaction}\label{nofeedback}


It is common to model dust in PPDs as passive particles when the dust-to-gas ratio $\epsilon \ll 1$ \citep[e.g.][]{stoll16,flock17}. Here we demonstrate that this condition is problem-dependent, and it alone may be insufficient to justify neglecting particle back-reaction. Specifically, we show that feedback has significant impact in weakly VSI-turbulent discs. 

\begin{figure}
\includegraphics[width=\linewidth]{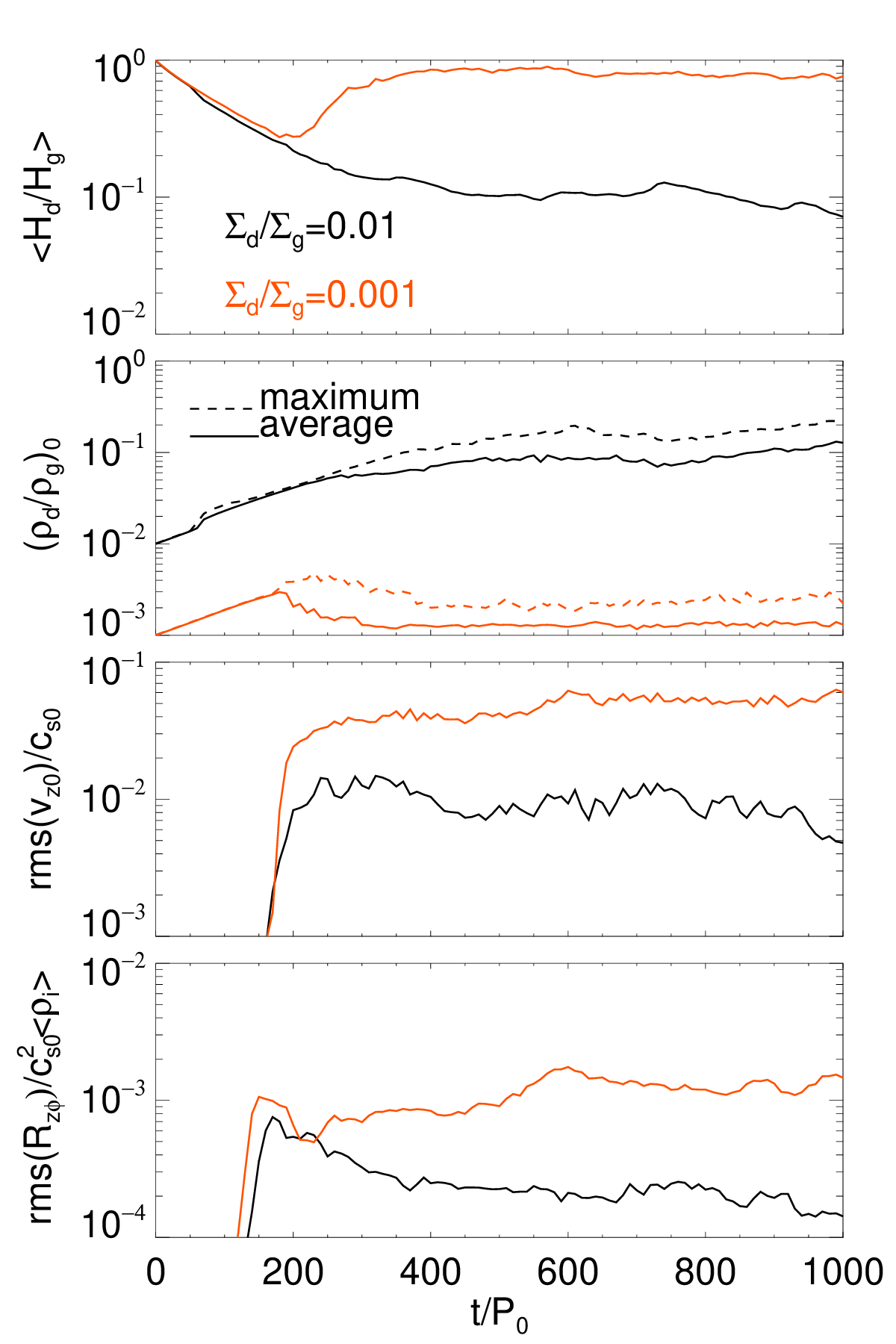}
\caption{{  Top to bottom: evolution of dust-scale height, midplane dust-to-gas ratios, midplane vertical velocity, and vertical turbulence parameter;} in discs with metallicity $Z=10^{-3}$ (orange) and $Z=10^{-2}$ (black, as considered in the main text). The radial temperature profile and particle size is fixed to $q=0.3$ and $\Tstop=10^{-3}$, respectively. 
\label{sub_solZ}}
\end{figure}

We consider a disc with $q=0.3$ and particles with $\Tstop=10^{-3}$. Fig. \ref{sub_solZ} compares the simulation with $Z=10^{-2}$, as presented in the main text, to a disc with $Z=10^{-3}$. They evolve similarly until $t\sim200P_0$, at which point dust in the $Z=10^{-3}$ disc undergoes complete remixing by the VSI; whereas dust continues to settle in the $Z=10^{-2}$ disc. In the latter case $\avg{\epsilon_0}\simeq 0.04\ll 1$ at $t=200P_0$, but this is apparently sufficient to overcome the VSI. 
{  As explained in \S\ref{buoyant_stabilisation}, this is because the VSI is sensitive to buoyancy forces induced by vertical gradients in the dust-to-gas ratio. Even an $\epsilon$ of order $\hgas\ll 1$ can provide effective stabilisation against the VSI.} 

\section{Kelvin-Helmholtz instabilities}\label{KHI}



Axisymmetric simulations preclude Kelvin-Helmholtz instabilities associated with the vertical shear between a dusty midplane and the dust-free gas above and below it \citep{chiang10,lee10}. Nevertheless, we can use the axisymmetric profiles reached in our simulations to examine stability against the KHI. In the literature this is usually assessed  through the Richardson number,  $\mathrm{Ri} = N_z^2/(\p_z v_\phi)^2$. Numerical simulations find KHI occurs for $\mathrm{Ri}\lesssim \mathrm{Ri}_*$, where the critical value  $\mathrm{Ri}_*$ depends on the dust layer parameters \citep{lee10}.  

\begin{figure}
\includegraphics[width=\linewidth]{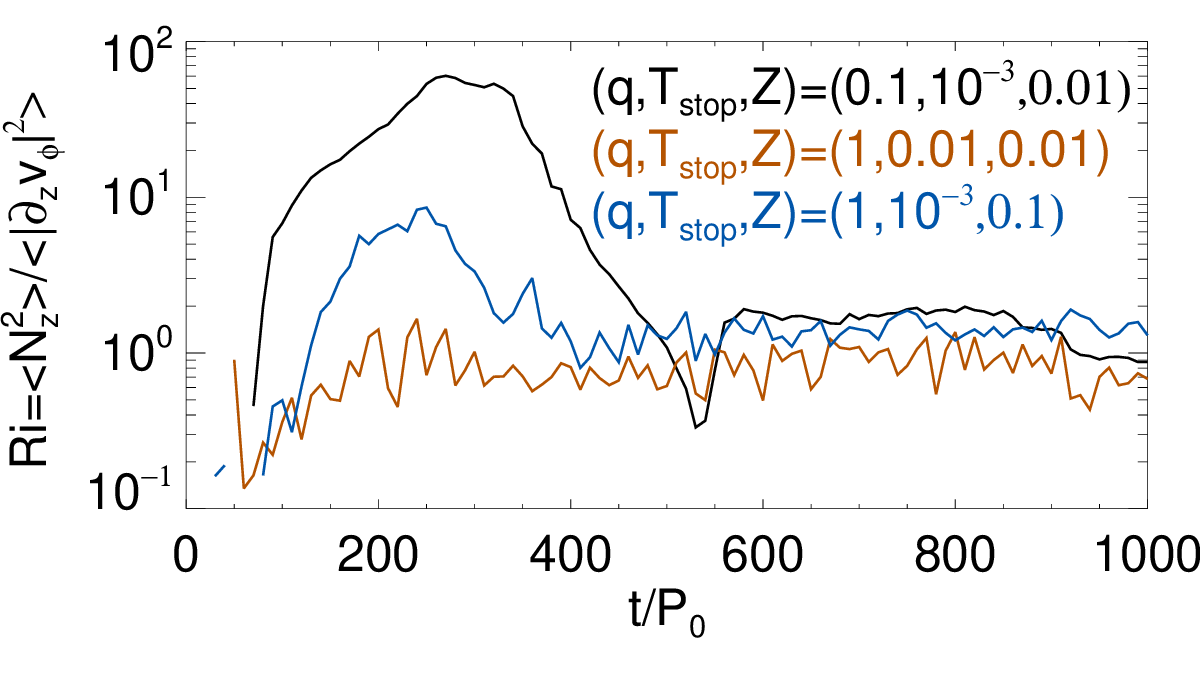}
\caption{Evolution of the Richardson number in the dust layer, for simulations where significant dust settling occurs. 
\label{richardson}}
\end{figure}

Fig. \ref{richardson} show the evolution of $\mathrm{Ri}$ in selected simulations where dust settles down to a few per cent of $\Hgas$. For  $Z=0.01$ \citeauthor{lee10} finds $\mathrm{Ri}_*\sim 0.1$, which is smaller than that the corresponding simulations. These cases are therefore stable against the KHI. 
However, extrapolating the numerical results of \citeauthor{lee10} gives $\mathrm{Ri}_*\simeq 2$ for $Z=0.1$, whereas we find $\mathrm{Ri}=1$---2 in this case. The $Z=0.1$ simulation may thus be marginally unstable to KHI. 
Full 3D simulations will be required to check this. 

\end{document}